\documentclass[a4paper,11pt]{article}
\usepackage{pos}

\def\l{\lambda}

\newcommand{\nc}{\newcommand}
\nc{\beq}{\begin{equation}}
\nc{\eeq}{\end{equation}}
\nc{\bea}{\begin{eqnarray}}
\nc{\eea}{\end{eqnarray}}
\def\IZ{\mathbb{Z}}

\def\ov{\overline}

\title{On Formulating the Non-Geometric Scalar Potentials%Non-geometric fluxes%Towards Non-geometric Flux Phenomenology
}
\author*[a]{George K. Leontaris}
\author[b]{Pramod Shukla}

\affiliation[a]{Physics Department, University of Ioannina\\
	University Campus, Ioannina 45110, Greece}

\affiliation[b]{Department of Physiccal Sciences, Bose Institute,\\
	Unified Academic Campus, EN 80, Sector V, Bidhannagar, Kolkata 700091, India}

\emailAdd{leonta@uoi.gr}
\emailAdd{pshukla@jcbose.ac.in}

\abstract{In the context of four-dimensional type II supergravities, the successive application of various S/T-dualities leads to a generalized notion of fluxes, which includes certain (non-)geometric fluxes along with the standard RR and NS-NS $p$-form fluxes. These fluxes induce a diverse set of superpotential couplings leading to scalar potentials with a very rich structure, which may possibly result in a vast landscape of physical vacua. However such scalar potentials typically consist of a huge number of terms and in order to make any attempt for phenomenological model building with some analytic understanding/control it is necessary to formulate them in some compact and concise form. Along these lines we review various equivalent methods of deriving the same scalar potential 
 through  a set of master formulae, which may open up a new avenue for model building  using non-geometric fluxes.}

\FullConference{Corfu Summer Institute 2023 "School and Workshops on Elementary Particle Physics and Gravity" (CORFU2023)\\
	23 April - 6 May , and 27 August - 1 October, 2023\\
	Corfu, Greece\\}

\begin{document}
	\maketitle

\section{Introduction}

\noindent
In the context of four-dimensional type IIB supergravities, the study of non-geometric fluxes has attracted a huge amount of interest in last two decades. The existence of such fluxes are supported through the presence of various dualities which, in general, have played a crucial role in connecting different limits of string theories,  providing supplemental interpretations and new perspectives. For instance, considering the NS-NS three-form flux $H_{mnp}$ of the type II supergravities, a chain of successive T-dualities leads to various (non-)geometric fluxes  \cite{Shelton:2005cf,Shelton:2006fd},
\bea
\label{eq:Tdual}
& & H_{mnp} \to \omega_{np}{}^m \to Q_p{}^{mn} \to R^{mnp},
\eea
where $\omega_{np}{}^m$ denotes the geometric fluxes (also known as metric flux) while $Q_p{}^{mn}$ and $R^{mnp}$ denote the set of non-geometric fluxes. For type IIB orientifold models one can choose the holomorphic involutions such that the geometric flux $\omega_{np}{}^m$ and the non-geometric flux $R^{mnp}$ are absent \cite{Benmachiche:2006df,Grana:2006hr,Aldazabal:2006up,Robbins:2007yv}, and one is left with a pair of S-dual flux RR and NS-NS three-form fluxes $(F, H)$, and the non-geometric $Q$-flux. However after adding these non-geometric fluxes, the underlying modular invariance in broken, and it can only be restored through the inclusion of a new kind of non-geometric flux, the so-called $P$-flux, which is S-dual to the non-geometric $Q$-flux ~\cite{Aldazabal:2006up,Font:2008vd,Guarino:2008ik, Hull:2004in, Aldazabal:2008zza, Kumar:1996zx, Hull:2003kr}. Subsequently if one demands a successive application of such S/T-dualities, one needs two additional S-dual pairs of fluxes, denoted as $(P', Q')$ and $(H',F')$ \cite{Aldazabal:2006up,Aldazabal:2008zza,Shukla:2016hyy,Lombardo:2016swq,Lombardo:2017yme}. The presence of such four S-dual pairs of  `generalized' fluxes, namely $(F, H), (Q, P), (P', Q')$ and $(H', F')$, induces a wide range of superpotential couplings which can subsequently result in a very rich structure for the four-dimensional effective scalar potential. However, how many and which type of fluxes can be simultaneously turned-on in a given concrete construction remains an open challenge for the most generic models. Some initiatives have been taken in this direction leading to a set of flux constraints arising in the form of Bianchi identities and tadpole cancellation conditions \cite{Aldazabal:2006up,Aldazabal:2008zza,Lombardo:2016swq,Lombardo:2017yme,Font:2008vd,Plauschinn:2021hkp,Leontaris:2023lfc}. 

The study of models based on non-geometric flux compactifications has been motivated along (but not limited to) the following points:
\begin{itemize}
\item 
One of the main attractive features of non-geometric flux models is the fact that one can in principle stabilize all types of moduli ~\cite{Derendinger:2004jn,Grana:2012rr,Dibitetto:2012rk, Danielsson:2012by, Blaback:2013ht, Damian:2013dq, Damian:2013dwa, Hassler:2014mla, Ihl:2007ah, deCarlos:2009qm, Danielsson:2009ff, Blaback:2015zra, Dibitetto:2011qs} without invoking non-perturbative effects in the superpotential, or utilizing any corrections of the K\"ahler potential. In this regard, it should be emphasized that this approach of stabilization includes also  the K\"ahler moduli which, in conventional flux compactifications, are protected by the so-called ``no-scale structure". From this point of view, the subsequent consideration of non-geometric fluxes  which can generically induce the superpotential couplings for the K\"ahler moduli as well, has emerged as an important ingredient in the area of moduli stabilization and generic model building~\cite{Derendinger:2004jn,Grana:2012rr,Dibitetto:2012rk, Danielsson:2012by, Blaback:2013ht, Damian:2013dq, Damian:2013dwa, Hassler:2014mla, Ihl:2007ah, deCarlos:2009qm, Danielsson:2009ff, Blaback:2015zra, Dibitetto:2011qs,Damian:2018tlf,CaboBizet:2019sku, Plauschinn:2018wbo, Damian:2023ote}.

\item
However, as we have earlier argued, although the inclusion of the new (non-geometric) fluxes may greatly facilitate the presence of new flux vacua, it is also true that it introduces numerous inevitable complexities due to the huge number of flux-induced terms in the scalar potential. For example, in the context of type IIB on ${\mathbb T}^6/({\mathbb Z}_2\times{\mathbb Z}_2)$ orientifold- it has been found that the resulting four-dimensional scalar potential is very often so huge that it gets hard to analytically solve the extremization conditions. Thus, one has to look either for a simplified Ansatz by switching off certain flux components at a time, or else one has to opt for a rather involved numerical analysis; for phenomenological model building attempts with (non-)geometric fluxes see ~\cite{Aldazabal:2008zza,Font:2008vd,Guarino:2008ik,Danielsson:2012by,Damian:2013dq, Damian:2013dwa, Hassler:2014mla, Blumenhagen:2015qda, Blumenhagen:2015kja,Blumenhagen:2015jva, Blumenhagen:2015xpa,  Li:2015taa, Blumenhagen:2015xpa,Shukla:2016xdy,Shukla:2022srx,Marchesano:2020uqz,Marchesano:2021gyv, Damian:2023ote, Prieto:2024shz}.

\item 
Apart from the phenomenological applications, understanding the higher dimensional origin of the four-dimensional scalar potential induced from the generalized flux superpotential has helped in exploiting the applications of symplectic geometries \cite{Ceresole:1995ca, Taylor:1999ii, Blumenhagen:2003vr, D'Auria:2007ay, Shukla:2015hpa, Blumenhagen:2015lta, Shukla:2016xdy, Leontaris:2023mmm}. 

\item 
The natural incorporation of non-geometric fluxes in the context of Double Field Theory and its subsequent connection with the non-geometric type II models arrived at after dimensional reduction has opened a completely new avenue to explore the higher dimensional origins of the generalized flux superpotentials \cite{Blumenhagen:2013hva,Blumenhagen:2015lta,Gao:2017gxk}.

\item 

In addition to the phenomenological model building motivations, the interesting connections among the ingredients of superstring flux-compactifications and those of the gauged supergravities have significant relevance in understanding both sectors as fluxes in one setting are related to the gauging in the other one  \cite{Derendinger:2004jn, Derendinger:2005ph, Shelton:2005cf, Aldazabal:2006up, Dall'Agata:2009gv, Aldazabal:2011yz, Aldazabal:2011nj,Geissbuhler:2011mx,Grana:2012rr,Dibitetto:2012rk, Villadoro:2005cu}.

\end{itemize}
  
\noindent
The main aim of this work is to present a brief review about the scalar potential formulations which is a crucial step towards performing any model building attempts using non-geometric setups. The F-term scalar potential governing the dynamics of the ${\cal N}=1$ low energy effective supergravity can be computed from the K\"ahler potential ($K$) and the flux induced holomorphic superpotential $(W)$ by considering the following well known formula,
\bea
\label{eq:Vtot}
& & \hskip-1cm V=e^{K}\Big(K^{{A} \ov{B}} \, (D_{A} W) \, (\ov D_{\ov {B}} \ov{W}) -3 |W|^2 \Big)  \,,\,
\eea
where the covariant derivatives for a chiral coordinate $A$ is defined as $D_A W = \partial_A W + (\partial_T K) \, W$. This general expression (\ref{eq:Vtot}) has resulted in a series of the so-called ``master-formulae" for the scalar potential for a given set of K\"ahler- and the super-potentials; e.g.~see \cite{Cicoli:2007xp,Shukla:2015hpa,Shukla:2016hyy,Cicoli:2017shd,Gao:2017gxk,Shukla:2019wfo,AbdusSalam:2020ywo,Cicoli:2021dhg,Cicoli:2021tzt,Marchesano:2020uqz,Leontaris:2022rzj,Leontaris:2023lfc, Leontaris:2023mmm, Biswas:2024ewk}
Along these lines we plan to briefly recall the iterative steps taken in the literature to understand the insights of the (generalized) flux superpotential via the understanding of the effective scalar potential. In this regard, the initial model building studies have been performed by considering the 4D effective potential derived by merely knowing the K\"ahler- and the super-potentials~\cite{Danielsson:2012by, Blaback:2013ht, Damian:2013dq, Damian:2013dwa, Blumenhagen:2013hva, Villadoro:2005cu, Robbins:2007yv, Ihl:2007ah, Gao:2015nra, Plauschinn:2021hkp}, and without having a complete understanding of their ten-dimensional origin. This issue has been taken into consideration in recent years via two approaches; the first one being through the Double Field Theory (DFT) \cite{Andriot:2013xca, Andriot:2011uh, Blumenhagen:2015lta} and the second approach being through the study of the underlying supergravity theories \cite{Villadoro:2005cu, Blumenhagen:2013hva, Gao:2015nra, Shukla:2015rua, Shukla:2015bca,Andriot:2012wx, Andriot:2012an,Andriot:2014qla,Blair:2014zba,Shukla:2015hpa}. Some of the timelines about exploring the 10D origin of the 4D effective potential can be recalled as below:
\begin{itemize}

\item{{\bf Step-0:} In the context of standard type IIB flux compactification with the usual NS-NS and RR fluxes, $H_3$ and $F_3$, the four-dimensional scalar potential induced by the flux superpotential \cite{Gukov:1999ya,Dasgupta:1999ss} has been compactly derived through the dimensional reduction of the 10D kinetic pieces \cite{Taylor:1999ii, Blumenhagen:2003vr}. In this case the flux superpotential takes the following schematic form:
\bea
& & W_{\rm FH} = p_1(U^i) + S\, p_2(U^i), 
\eea
where $p_1, p_2$ are some cubic polynomials in terms of $U^i$ moduli. This leads to the so-called ``no-scale-structure" as there is no dependence on the K\"ahler moduli in $V$ defined in (\ref{eq:Vtot}), except from a ${\cal V}^{-2}$ dependence arising from the overall factor $e^K$. 

}
	
\item{{\bf Step-1:} Motivated by the study of 4D effective scalar potential in a type IIA flux compactification setup with geometric flux \cite{Villadoro:2005cu}, a rearrangement of the scalar potential induced through a generalized flux superpotential with non-geometric $Q$-fluxes on top of having the standard $H_3/F_3$ within a type IIB non-geometric framework, has been presented in \cite{Blumenhagen:2013hva}. This ``rearranged" scalar potential has a ``suitable" form which helps in anticipating the 10D origin of the 4D pieces, a process called as ``dimensional oxidation" of the non-geometric flux superpotential \cite{Blumenhagen:2013hva}.

In this case the flux superpotential takes the following schematic form:
\bea
& & W_{\rm FHQ} = p_1(U^i, T_\alpha) + S\, p_2(U^i), 
\eea
where $p_1, p_2$ are some cubic polynomials in terms of $U^i$ moduli. In addition, the polynomial $p_1$ has a linear dependence on the K\"ahler moduli $T_\alpha$ which breaks the no-scale structure and creates the possibilities of stabilizing all moduli without including any (non-)perturbative effects from string-loops or instanton series. 

}

\item{{\bf Step-2:} In order to restore the S-duality invariance broken by including the non-geometric $Q$-flux in the type IIB ${\mathbb T}^6/{\left({\mathbb Z}_2 \times {\mathbb Z}_2\right)}$-orientifold setup, the proposal of \cite{Blumenhagen:2013hva} was further generalized in \cite{Gao:2015nra} via the inclusion of the so-called $P$-flux which is S-dual to the $Q$-flux. In the meantime the prescription was further extended for the odd axion models within a type IIB compactification on ${\mathbb T}^6/{{\mathbb Z}_4}$-orientifold in \cite{Shukla:2015bca,Shukla:2015rua}.

In this case the flux superpotential takes the following schematic form:
\bea
& & W_{\rm FHQP} = p_1(U^i, T_\alpha) + S\, p_2(U^i, T_\alpha), 
\eea
where $p_1, p_2$ are some cubic polynomials in terms of $U^i$ moduli whereas they both have only a linear dependence on the K\"ahler moduli $T_\alpha$. The scalar potential induced through this superpotential beyond the toroidal case has been studied in \cite{Shukla:2016hyy}.

}

\item{{\bf Step-3:} One can further extend the S/T dual completion arguments to arrive at a more general flux superpotential of the following form, via including two more S-dual pairs of fluxes, usually denoted as $(P', Q')$ and $(H', F')$ \cite{Aldazabal:2006up, Aldazabal:2008zza, Aldazabal:2010ef, Lombardo:2016swq, Lombardo:2017yme}
\bea
& & W_{\rm FHQPP'Q'H'F'} = p_1(U^i, T_\alpha) + S\, p_2(U^i, T_\alpha), 
\eea
where now $p_1, p_2$ are some cubic polynomials in terms of $U^i$ as well as $T_\alpha$ moduli. Let us note that such a U-dual completed flux superpotential has been explicitly known since quite some time, and for the toroidal type IIB $\mathbb T^6/(\mathbb Z_2 \times \mathbb Z_2)$ orientifold based model it has a total of 128 flux parameters and 7 complexified variable \cite{Aldazabal:2006up}. However its insights (or any phenomenological application in model building) have not been explored in detail because of the huge size of the scalar potential which was recently found to have a total of 76276 terms \cite{Leontaris:2023lfc,Leontaris:2023mmm}! 

}

\end{itemize}

 \noindent
In the current work we plan to briefly elaborate on these iterative steps taken to understand the deeper insights of the four-dimensional scalar potentials in the presence of non-geometric fluxes. For this task we basically follow two approaches: (i).~To formulate the scalar potential using the metric of the internal toroidal background \cite{Blumenhagen:2013hva, Gao:2015nra, Shukla:2015rua}, and (ii).~To formulate the scalar potential using the symplectic ingredients \cite{Shukla:2015hpa,Blumenhagen:2015lta,Shukla:2016hyy,Shukla:2019wfo,Shukla:2019dqd,Shukla:2019akv, Leontaris:2023lfc,Leontaris:2023mmm,Biswas:2024ewk}. These beyond toroidal formulations are generically valid for arbitrary number of complex structure moduli as well as K\"ahler moduli, and on top of it, do not need the knowledge of internal background metric.

%%%%%%%%%%%%%%%%%%%%%%%%%%%%%%%%%%%%%%%%%%%%%%%%%%%%%%%%%%%%%%%%%%%%%%%%%%%%%%%%%%%%%%%%%%
%%%%%%%%%%%%%%%%%%%%%%%%%%%%%%%%%%%%%%%%%%%%%%%%%%%%%%%%%%%%%%%%%%%%%%%%%%%%%%%%%%%%%%%%%%

\section{Internal metric formulation}
\label{sec_standard}

Most of the studies related to the 4D type II effective theories in a non-geometric flux compactification framework have been centered around toroidal examples, and in particular using a ${\mathbb T}^6/({\mathbb Z}_2 \times {\mathbb Z}_2)$ orientifold. An obvious reason is its relatively simpler structure to perform explicit and concrete computations, which have led toroidal setups to serve as promising toolkit in studying concrete examples. In this regard, one useful ingredient is the metric of the toroidal background which is explicitly known unlike the case of generic CY backgrounds, and we will demonstrate the use of internal metric in writing down the scalar potential pieces a la `dimensional oxidation' approach of \cite{Blumenhagen:2013hva, Gao:2015nra, Shukla:2015bca, Leontaris:2023lfc}.

%%%%%%%%%%%%%%%%%%%%%%%%%%%%%%%%%%%%%%%%%%%%%%%%%%%%%%%%%%%%%%%%%%%%%%%%%%%%%%%%%%%%%%%%%%
%%%%%%%%%%%%%%%%%%%%%%%%%%%%%%%%%%%%%%%%%%%%%%%%%%%%%%%%%%%%%%%%%%%%%%%%%%%%%%%%%%%%%%%%%%

\subsection{Type IIB model using a ${\mathbb T}^6 / \left(\mathbb Z_2\times \mathbb Z_2\right)$ orientifold}

Let us start by briefly revisiting the relevant features of a concrete setup in the framework of the type IIB orientifold compactification using the well studied ${\mathbb T}^6 / \left(\mathbb Z_2\times \mathbb Z_2\right)$ orbifold, where the two $\mathbb Z_2$ actions are defined as,
\bea
\label{thetaactions}
& & \theta:(z^1,z^2,z^3)\to (-z^1,-z^2,z^3)\\
 & &   \ov\theta:(z^1,z^2,z^3)\to (z^1,-z^2,-z^3)\, . \nonumber
 \eea
Next, the orientifold action is defined via ${\cal O} \equiv \Omega_p\,  I_6 \, (-1)^{F_L}$ where $\Omega_p$ is the worldsheet parity, $F_L$ is left-fermion number and $I_6$ denotes the holomorphic involution defined as,
\bea
\label{eq:orientifold}
& & I_6 : (z^1,z^2,z^3)\rightarrow (-z^1,-z^2,-z^3)\,,
\eea
which subsequently results in a setup of $O3/O7$-type. The complexified coordinates ($z^i$) on the six-torus ${\mathbb T}^6={\mathbb T}^2\times {\mathbb T}^2\times {\mathbb T}^2$ are defined as below,
\bea
z^1=x^1+ U^1  x^2, ~ z^2=x^3+ U^2 x^4,~ z^3=x^5+ U^3 x^6 ,
\eea
where the three complex structure moduli $U^i$'s can be written as
$U^i= v^i - i\, u^i,\,\,i=1,2,3$. Now, the holomorphic three-form $\Omega_3=dz^1\wedge dz^2\wedge dz^3$ can be expanded as,
\bea
\label{eq:Omega3}
& & \hskip-0.5cm  \Omega_3\, = \alpha_0 +  \, U^1 \, \alpha_1 + U^2 \, \alpha_2 + U^3 \alpha_3 \\
& & \hskip0.5cm  +\, U^1 \, U^2 \,  U^3 \, \beta^0 -U^2 \, U^3 \, \beta^1- U^1 \, U^3 \, \beta^2 - U^1\, U^2 \, \beta^3 \,, \nonumber
\eea
where we have chosen the following basis of the closed three-forms,
\bea
\label{formbasis}
& & \hskip-1cm \alpha_0=1\wedge 3\wedge 5\,, \, \alpha_1=2\wedge 3\wedge 5\, , \,  \alpha_2=1\wedge 4\wedge 5\, , \, \alpha_3=1\wedge 3\wedge 6, \\
& & \hskip-1cm \beta^0= 2\wedge 4\wedge 6,\, \beta^1= -1\wedge 4\wedge 6,\,  \beta^2=-2\wedge 3\wedge 6,\, \beta^3=-\, 2\wedge 4\wedge 5\,.\nonumber
\eea
In the above we use the shorthand notations such as $1\wedge 3\wedge 5 = dx^1\wedge dx^3\wedge dx^5$ etc. along with the normalization $\int \alpha_\Lambda \wedge \beta^\Delta=-\delta_\Lambda{}^\Delta$. Using these ingredients, the holomorphic three-form can also be expressed in terms of the symplectic period vectors $({\cal X}^\Lambda, {\cal F}_\Lambda)$ as $\Omega_3\,  \equiv {\cal X}^\Lambda \alpha_\Lambda - {\cal F}_\Lambda \beta^\Lambda$, where the complex structure moduli dependent prepotential ${\cal F}$ is given as,
\bea
\label{eq:prepotentialA}
& & {\cal F} = \frac{{\cal X}^1 \, {\cal X}^2 \, {\cal X}^3}{{\cal X}^0} = U^1 \, U^2 \, U^3
\eea 
which results in the following period-vectors,
\bea
& & {\cal X}^0=1\,, \quad   {\cal X}^1=U^1\, , \quad {\cal X}^2=U^2\, , \quad  {\cal X}^3=U^3\, , \\
& & {\cal F}_0=\, -\, \, U^1 \, U^2 \, U^3\, , \quad  {\cal F}_1= U^2 \, U^3\, , \quad  {\cal F}_2=U^3 \, U^1\, , \quad  {\cal F}_3=U^1 \, U^2\,.  \nonumber
\eea
Now, using the same shorthand notations we choose the following bases for the orientifold even two-forms $\mu_\alpha$, and their dual four-forms ${\tilde\mu^\alpha}$,
\bea
& & \hskip-0.7cm \mu_1 = 1 \wedge 2,  \quad \mu_2 = 3 \wedge 4, \quad  \mu_3 = 5 \wedge 6; \\
& & \hskip-0.7cm \tilde{\mu}^1 = 3 \wedge 4 \wedge 5 \wedge 6,  \quad \tilde{\mu}^2 = 1 \wedge 2 \wedge 5 \wedge 6, \quad  \tilde{\mu}^3 = 1 \wedge 2 \wedge 3 \wedge 4, \nonumber
\eea
Let us mention that for this toroidal orientifold construction there are no two-forms which are anti-invariant under the orientifold projection, i.e. $h^{1,1}_-(X) = 0$, and similarly there dual four-forms are also trivial, and therefore no ${B}_2$ and $C_2$ moduli as well as no geometric-flux components will be present in this model; for the construction of concrete type IIB orientifold models with odd moduli, e.g. see \cite{Gao:2013pra,Carta:2020ohw,Altman:2021pyc, Carta:2022web, Crino:2022zjk, Shukla:2022dhz, Cicoli:2021tzt}.

The other chiral variables are the so-called axio-dilaton $S$ and the complexified K\"ahler moduli which are defined as,
\bea
\label{N=1-coordinates}
& & S= C_{0} + i\, e^{-\phi}\, \qquad {\cal J}  = C^{(4)} -  \frac{i}{2} \, J \wedge J \equiv T_\alpha \, \tilde\mu^\alpha\,,
\eea
where $J = t^\alpha \mu_\alpha$ is the K\"ahler form involving the (Einstein-frame) two-cycle volume moduli $t^\alpha$ while the moduli $T_\alpha = \rho_\alpha - i \, \tau_\alpha$ consist of RR axions $C^{(4)}_{ijkl}$ and the four-cycle volume moduli $\tau_{ijkl}$ which in terms of six-dimensional components are given as follows,
\bea
T_1 = C^{(4)}_{3456} - i \, \tau_{3456}, \qquad T_2 = C^{(4)}_{1256} - i \, \tau_{1256}, \qquad T_3 = C^{(4)}_{1234} - i \, \tau_{1234},
\eea
where $\tau_1 = \, t^2 \, t^3, \,\, \tau_2 = \, t^3 \, t^1, \,\, \tau^3 =  \, t^1 \, t^2$ are expressed in the Einstein-frame. The overall volume (${\cal V}$) of the sixfold (in the Einstein-frame) can be given as,
\bea
\label{eq:vol}
& & {\cal V} = t^1 \, t^2 \, t^3 =  \sqrt{\tau_1 \tau_2 \tau_3}, \qquad \tau_\alpha = \frac{\partial {\cal V}}{\partial t^\alpha},
\eea
where a useful relation between the two-cycle volumes $t^\alpha$ and the four-cycles volumes $\tau_\alpha$ can be given as below,
\bea
\label{twoinfour}
& & t^1=\sqrt{\tau_2\,  \tau_3\over \tau_1}\,, \qquad  t^2=\sqrt{\tau_1\,  \tau_3\over \tau_2}\,, \quad t^3=\sqrt{\tau_1\,  \tau_2\over \tau_3}\,.
\eea
Another crucially relevant ingredient in our current study is the information about the internal metric of the toroidal sixfold. %The non-vanishing components of the metric in string-frame are
%\bea
%\label{eq:gMN}
%    g_{MN}={\rm blockdiag}\Big({e^{\phi\over 2}\over {\sqrt{\tau_1\, \tau_2
%          \, \tau_3}}} \, \, \tilde g_{\mu\nu}, \, \, g_{ij}\Big) \, .
%\eea
It turns out that the internal metric $g_{ij}$ is block-diagonal and
has the following non-vanishing components,
\bea
\label{eq:gij-ts}
& & \hskip-0.8cm g_{11}=\frac{t^1}{u^1}\,,  \qquad  g_{12}=\frac{t^1 v^1}{u^1} =g_{21}\, , \qquad
g_{22}=\frac{t^1((u^1)^2+(v^1)^2)}{u^1}\, ,\nonumber\\
& & \hskip-0.8cm g_{33}=\frac{t^2}{u^2}\, , \qquad   g_{34}=\frac{t^2 v^2}{u^2}=g_{43}\, , \qquad
g_{44}=\frac{t^2((u^2)^2+(v^2)^2)}{u^2}\, ,\\
& & \hskip-0.8cm g_{55}=\frac{t^3}{u^3}\,, \qquad   g_{56}=\frac{t^3 v^3}{u^3}=g_{65}\,, \qquad
g_{66}=\frac{t^3((u^3)^2+(v^3)^2)}{u^3}\, .\nonumber
\eea
These internal metric components can be written out in a more suitable form, to be utilized later, by using the four-cycle volumes $\tau_i$'s and the same is given as below,
\bea
\label{eq:gij-taus}
& & \hskip-0.8cm g_{11}=\frac{\sqrt{\tau_2} \,  \sqrt{\tau_3}}{u^1 \, \sqrt{\tau_1}}\,,   \quad g_{12}=\frac{v^1 \, \sqrt{\tau_2} \,  \sqrt{\tau_3}}{ u^1 \, \sqrt{\tau_1}} =g_{21}\, , \quad
g_{22}= \frac{\left((u^1)^2 + (v^1)^2\right)\sqrt{\tau_2} \,  \sqrt{\tau_3}}{u^1 \, \sqrt{\tau_1}} ,\nonumber\\
& & \hskip-0.8cm g_{33}=\frac{\sqrt{\tau_1} \,  \sqrt{\tau_3}}{u^2 \, \sqrt{\tau_2}}\,,   \quad g_{34}=\frac{v^2 \, \sqrt{\tau_1} \,  \sqrt{\tau_3}}{u^2 \, \sqrt{\tau_2}} =g_{43}\, , \quad
g_{44}= \frac{\left((u^2)^2 + (v^2)^2\right)\sqrt{\tau_1} \,  \sqrt{\tau_3}}{u^2 \, \sqrt{\tau_2}} ,\\
& & \hskip-0.8cm g_{55}=\frac{\sqrt{\tau_1} \,  \sqrt{\tau_2}}{u^3 \, \sqrt{\tau_3}}\,,   \quad g_{56}=\frac{v^3 \, \sqrt{\tau_1} \,  \sqrt{\tau_2}}{u^3 \, \sqrt{\tau_3}} =g_{65}\, , \quad
g_{66}= \frac{\left((u^3)^2 + (v^3)^2\right)\sqrt{\tau_1} \,  \sqrt{\tau_2}}{u^3 \, \sqrt{\tau_3}}. \nonumber
\eea
For the current toroidal setup, the (tree level) K\"ahler potential takes the following form in terms of the $S$, $T$ and $U$ moduli,
\bea
\label{eq:K}
& & \hskip-1.6cm K = -\ln\left(-i(S-\ov{S})\right) -\sum_{j=1}^{3} \ln\left(i(U^j - \ov U^j)\right) - \sum_{\alpha=1}^{3} \ln\left(\frac{i\,(T_\alpha - \ov T_\alpha)}{2}\right).
\eea

%%%%%%%%%%%%%%%%%%%%%%%%%%%%%%%%%%%%%%%%%%%%%%%%%%%%%%%%%%%%%%%%%%%%%%%%%%%%%%%%%%%%%%%%%%
\subsection{U-dual completion of the flux Superpotential}
%%%%%%%%%%%%%%%%%%%%%%%%%%%%%%%%%%%%%%%%%%%%%%%%%%%%%%%%%%%%%%%%%%%%%%%%%%%%%%%%%%%%%%%%%%

The four-dimensional effective scalar potential generically has an S-duality invariance following from the underlying ten-dimensional type IIB supergravity. This corresponds to the following $SL(2, \mathbb{Z})$ transformation,
\bea
\label{eq:SL2Za}
& & \hskip-1.5cm S \to \frac{a \, S+ b}{c \, S + d}\, \quad \quad {\rm where} \quad a d- b c = 1\,;\quad a,\ b,\ c,\ d\in \mathbb{Z}
\eea
Under this $SL(2, \mathbb{Z})$ transformation, the complex-structure moduli ($U^i$) and the Einstein-frame internal volume (${\cal V}$) are invariant. Moreover, the Einstein-frame chiral coordinate $T_{\alpha}$ is S-duality invariant, without orientifold odd axions, i.e. $h^{11}_-(X_6/{\cal O}) = 0$ \cite{Grimm:2007xm}. Subsequently, it turns out that the tree level K\"ahler potential given in eqn. (\ref{eq:K}) transforms as:
\bea
\label{eq:modularK}
e^K \longrightarrow |c \, S + d|^2 \, e^K\,.
\eea
This subsequently implies that the S-duality invariance of the physical quantities (such as gravitino mass-square $m_{3/2}^2 \propto e^K |W|^2$) suggests that the holomorphic superpotential, $W$ should have a modularity of weight $-1$, which means the following \cite{Font:1990gx, Cvetic:1991qm, Grimm:2007xm}
\bea
\label{eq:modularW}
& & W \to \frac{W}{c \, S + d}.
\eea
As we will discuss in the upcoming sections, a generic holomorphic superpotential, respecting the modular weight being -1, can have four S-dual pairs of fluxes denoted as $(F, H), \, (Q, P), \, (P', Q')$ and $(H', F')$ \cite{Aldazabal:2006up, Aldazabal:2008zza, Aldazabal:2010ef, Lombardo:2016swq, Lombardo:2017yme}. This set of eight fluxes transforms in the following manner under the $SL(2,{\mathbb Z})$ transformations,
\bea
\label{eq:modularFlux}
& & \hskip-1.5cm \left(\begin{array}{c} F \\ H \end{array}\right) \to \left(\begin{array}{cc} a  & \quad b \\ c & \quad d \end{array}\right)
\left(\begin{array}{c} F \\ H \end{array}\right)\,, \qquad \left(\begin{array}{c} Q \\ P \end{array}\right) \to \left(\begin{array}{cc} a  & \quad b \\ c & \quad d \end{array}\right)
\left(\begin{array}{c} Q \\ P \end{array}\right)\,, \\
& & \hskip-1.5cm \left(\begin{array}{c} H^\prime \\ F^\prime \end{array}\right) \to \left(\begin{array}{cc} a  & \quad b \\ c & \quad d \end{array}\right)
\left(\begin{array}{c} H^\prime \\ F^\prime \end{array}\right)\,, \qquad \left(\begin{array}{c} P^\prime \\ Q^\prime \end{array}\right) \to \left(\begin{array}{cc} a  & \quad b \\ c & \quad d \end{array}\right) \left(\begin{array}{c} P^\prime \\ Q^\prime \end{array}\right). \nonumber
\eea
Under the $SL(2, \IZ)$ transformations, the various fluxes can readjust themselves to respect the modularity condition (\ref{eq:modularW}) in the following two ways \cite{Aldazabal:2006up},
\bea
& & (i). \quad S \to S + 1, \qquad (ii). \quad S \to -\frac{1}{S}.
\eea
Note that the first case simply corresponds to a shift in the universal axion $C_0 \to C_0 + 1$ which amounts to having a constant rescaling of the K\"ahler potential as $e^K \to |d|^2\, e^K$,  and the superpotential as $W \to W/d$. This follows from Eqs.~(\ref{eq:modularK})-(\ref{eq:modularW}) due to the fact that $S \to S +1$ simply corresponds to $c=0$ case in the $SL(2, \mathbb{Z})$ transformation (\ref{eq:SL2Za}). The second case is quite peculiar in the sense that it corresponds to the following transformation of the universal axions and the dilaton,
\bea
\label{eq:modularS}
& & C_0 \to -\frac{C_0}{s^2 + C_0^2}, \qquad s \to \frac{s}{s^2 + C_0^2},
\eea
which takes $g_s \to g_s^{-1}$ and hence is known as strong-week duality or S-duality. This relation (\ref{eq:modularS}) shows that $C_0/s$ flips sign under $S$-duality, something which will be useful in understanding the modular completion of the scalar potential later on. From now onwards we will focus only on the second case, i.e. on strong/weak duality. This means that under the $SL(2,{\mathbb Z})$ transformation of the second type which simply takes the axio-dilaton $S \to - 1/S$, the fluxes can be considered to transform as below,
\bea
\label{eq:modularFlux-simp}
& & {H} \to {F}, \quad {F} \to -{H}, \quad {Q} \to - {P}, \quad {P} \to {Q},\\
& & {F'} \to {H'}, \quad {H'} \to -{F'}, \quad {P'} \to - {Q'}, \quad {Q'} \to {P'}. \nonumber
\eea
Now let us mention that the complete set of fluxes, including the so-called prime fluxes $P', Q', H'$ and $F'$ which are some mixed-tensor quantities, have the following index structure \cite{Aldazabal:2006up,Aldazabal:2008zza,Aldazabal:2010ef,Lombardo:2016swq,Lombardo:2017yme},
\bea
\label{eq:All-flux}
& & F_{ijk}, \qquad H_{ijk}, \qquad Q_i{}^{jk}, \qquad P_i{}^{jk}, \\
& & P'^{i,jklm}, \quad Q'^{i,jklm}, \quad H'^{ijk,lmnpqr}, \quad F'^{ijk,lmnpqr}. \nonumber
\eea
Subsequently, one can understand $(P', Q')$ flux as a $(1,4)$ tensor such that only the last four-indices are anti-symmetrized, while $(H', F')$ flux can be considered as a $(3, 6)$ tensor where first three indices and last six indices are separately anti-symmetrized. These can also be understood as
\bea
\label{eq:shortPrimed-flux}
& & P'_{ij}{}^k = \frac{1}{4!} \,  \epsilon_{ijlmnp}\, P'^{k,lmnp}, \qquad \qquad \, \,  Q'_{ij}{}^k = \frac{1}{4!} \,  \epsilon_{ijlmnp}\, Q'^{k,lmnp}, \\
& & H'^{ijk} = \frac{1}{6!} \,  \epsilon_{lmnpqr}\, H'^{ijk,lmnpqr}, \qquad F'^{ijk} = \frac{1}{6!} \,  \epsilon_{lmnpqr}\, F'^{ijk,lmnpqr}. \nonumber
\eea
Using (\ref{eq:shortPrimed-flux}) for prime fluxes, the number of flux parameters consistent with our toroidal orientifold for $(P', Q')$ are 24 each, while those of $(H', F')$ are 8 each. Subsequently, one has a total of 8+8+24+24+24+24+8+8 = 128 flux parameters allowed by the orientifold action, however these flux parameters are not all independent, since 
they are subject to restrictions derived from the Bianchi identities and tadpole cancellation conditions \cite{Ihl:2006pp,Ihl:2007ah,Robbins:2007yv,Shukla:2016xdy,Gao:2018ayp}. 
Using generalized geometry motivated through toroidal constructions, it has been argued that the type IIB superpotential governing the dynamics of the four-dimensional effective theory which respects the invariance under $SL(2, {\mathbb Z})^7$ symmetry can be given as \cite{Aldazabal:2006up,Aldazabal:2008zza,Aldazabal:2010ef,Lombardo:2016swq,Lombardo:2017yme},
\bea
\label{eq:W-all-flux}
& & W = \int_{X_3} \, \left(f_+ \, - \, S\, f_- \right) \, \cdot e^{{\cal J}} \wedge \Omega_3 \,,
\eea
where ${\cal J}$ denotes the complexified four-form defined in Eq.~(\ref{N=1-coordinates}), and one has the following expansions for the quantities $f_\pm$,
\bea
\label{eq:WIIBGENnew2b}
& & f_+  \cdot e^{{\cal J}} %= {\cal D}_1 \cdot e^{{\cal J}_c} 
= F + Q \triangleright {\cal J} +  P^\prime \diamond {\cal J}^2 + H^\prime \odot {\cal J}^3 \, \\
& & f_- \cdot e^{{\cal J}} %= {\cal D}_2 \cdot e^{{\cal J}_c} 
= H + P \triangleright {\cal J} +  Q^\prime \diamond {\cal J}^2 + F^\prime \odot {\cal J}^3 \,.\nonumber
\eea
Here, the flux-actions on the ${\cal J}_{ijkl}$ four-form polynomial pieces resulting in three-forms are \cite{Aldazabal:2006up,Aldazabal:2008zza,Aldazabal:2010ef,Lombardo:2016swq,Lombardo:2017yme}
\bea
\label{eq:fluxactionsIIB-udual-old}
&& \left(Q\triangleright {\cal J} \right)_{a_1a_2a_3} = \frac{3}{2} \, Q^{b_1b_2}_{[\underline{a_1}} \, {\cal J}_{\underline{a_2} \, \underline{a_3}] b_1 b_2} \, ,\\
&& \left(P\triangleright {\cal J} \right)_{a_1a_2a_3} = \frac{3}{2} \, P^{b_1b_2}_{[\underline{a_1}} \, {\cal J}_{\underline{a_2} \, \underline{a_3}] b_1 b_2} \, ,\nonumber\\
& & \nonumber\\
&& \left(P^\prime \diamond {\cal J}^2 \right)_{a_1a_2a_3} = \frac{1}{4} \, {P^\prime}^{c,b_1b_2b_3b_4}  \, {\cal J}_{[\underline{a_1} \underline{a_2} |c b_1|}\, {\cal J}_{\underline{a_3}] b_2 b_3 b_4}\,, \nonumber\\
&& \left(Q^\prime \diamond {\cal J}^2 \right)_{a_1a_2a_3} = \frac{1}{4} \, {Q^\prime}^{c,b_1b_2b_3b_4}  \, {\cal J}_{[\underline{a_1} \underline{a_2} |c b_1|}\, {\cal J}_{\underline{a_3}] b_2 b_3 b_4}\,, \nonumber\\
& & \nonumber\\
&& \left(H^\prime \odot {\cal J}^3 \right)_{a_1a_2a_3} = \frac{1}{192} \, {H^\prime}^{c_1 c_2 c_3, b_1 b_2b_3b_4b_5b_6}  \, {\cal J}_{[\underline{a_1}\underline{a_2} | c_1 c_2|}\, {\cal J}_{\underline{a_3}]c_3b_1b_2} \, {\cal J}_{b_3 b_4 b_5 b_6}\,, \nonumber\\
&& \left(F^\prime \odot {\cal J}^3 \right)_{a_1a_2a_3} = \frac{1}{192} \, {F^\prime}^{c_1 c_2 c_3, b_1 b_2b_3b_4b_5b_6}  \, {\cal J}_{[\underline{a_1}\underline{a_2} | c_1 c_2|}\, {\cal J}_{\underline{a_3}]c_3b_1b_2} \, {\cal J}_{b_3 b_4 b_5 b_6}\,. \nonumber
\eea
Subsequently one finds an explicit and expanded version of the generalized flux superpotential $W$ with 128 terms, each having one of the 128 flux parameters such that they are coupled with the complexified moduli resulting in cubic polynomial in $T_\alpha$ as well as $U^i$ moduli while being linear in the axio-dilaton $S$.

\subsection{Axionic fluxes and the scalar potential taxonomy}

%%%%%%%%%%%%%%%%%%%%%%%%%%%%%%%%%%%%%%%%%%%%%%%%%%%%%%%%%%%%%%%%%%%%%%%%%%%%%%%%%%%%%%%%%%
From the iterative models studied/revisited so far, one has the educated guess to invoke the following set of so-called axionic-flux combinations which will turn out to be extremely useful for rearranging the scalar potential pieces into a compact form \cite{Leontaris:2023lfc}, 
\bea
\label{eq:AxionicFlux-standardflux}
& & \hskip-0cm {\mathbb F}_{ijk} = \biggl({F}_{ijk} +\frac{3}{2} \, Q^{b_1b_2}_{[\underline{i}} \,\rho_{\underline{j} \, \underline{k}] b_1 b_2} + \frac{1}{4} \, {P^\prime}^{c,b_1b_2b_3b_4}  \, \rho_{[\underline{i} \, \underline{j} |c b_1|}\, \rho_{\underline{k}] b_2 b_3 b_4} \\
& & \hskip0.75cm +  \frac{1}{192} \, {H^\prime}^{c_1 c_2 c_3, b_1 b_2b_3b_4b_5b_6}  \, \rho_{[\underline{i}\, \underline{j} | c_1 c_2|}\, \rho_{\underline{k}]c_3b_1b_2} \, \rho_{b_3 b_4 b_5 b_6} \biggr) - C_0\, {\mathbb H}_{ijk}, \nonumber\\
& & \hskip-0cm {\mathbb H}_{ijk} = \biggl({H}_{ijk} +\frac{3}{2} \, P^{b_1b_2}_{[\underline{i}} \,\rho_{\underline{j} \, \underline{k}] b_1 b_2} + \frac{1}{4} \, {Q^\prime}^{c,b_1b_2b_3b_4}  \, \rho_{[\underline{i} \, \underline{j} |c b_1|}\, \rho_{\underline{k}] b_2 b_3 b_4} \nonumber\\
& & \hskip0.75cm +  \frac{1}{192} \, {F^\prime}^{c_1 c_2 c_3, b_1 b_2b_3b_4b_5b_6}  \, \rho_{[\underline{i}\, \underline{j} | c_1 c_2|}\, \rho_{\underline{k}]c_3b_1b_2} \, \rho_{b_3 b_4 b_5 b_6} \biggr), \nonumber\\
&& \nonumber\\
& & \hskip-0cm {\mathbb Q}_i{}^{jk} = \left({Q}_i{}^{jk} - \frac{1}{2} \, {P^\prime}^{c,jk b_1b_2}  \, \rho_{i \,c \, b_1 b_2}\, + \frac{1}{48} {H^\prime}^{jkc, b_1 b_2b_3b_4b_5b_6}  \, \rho_{i \, c \, b_1b_2} \, \rho_{b_3 b_4 b_5 b_6} \right) - C_0\,{\mathbb P}_i{}^{jk}, \nonumber\\
& & \hskip-0cm {\mathbb P}_i{}^{jk} = \left({P}_i{}^{jk} - \frac{1}{2} \, {Q^\prime}^{c,jk b_1b_2}  \, \rho_{i \,c \, b_1 b_2}\, + \frac{1}{48} {F^\prime}^{jkc, b_1 b_2b_3b_4b_5b_6}  \, \rho_{i \, c \, b_1b_2} \, \rho_{b_3 b_4 b_5 b_6} \right) , \nonumber\\
&& \nonumber\\
& & \hskip-0cm {\mathbb P'}^{i,jklm} = \left({P'}^{i,jklm}+ \frac{1}{4} \frac{}{} {H'}^{ij'k',l'm'jklm} \, \rho_{j'k'l'm'} \right) - C_0\, {\mathbb Q'}^{i,jklm}, \nonumber\\
& & \hskip-0cm {\mathbb Q'}^{i,jklm} = \left({Q'}^{i,jklm} + \frac{1}{4} \frac{}{} {F'}^{ij'k',l'm'jklm} \, \rho_{j'k'l'm'}\right), \nonumber\\
&& \nonumber\\
& & \hskip-0cm {\mathbb H'}^{ijk,lmnpqr} = {H'}^{ijk,lmnpqr} - C_0\, {\mathbb F'}^{ijk,lmnpqr}, \nonumber\\
& & \hskip-0cm {\mathbb F'}^{ijk,lmnpqr} = {F'}^{ijk,lmnpqr}. \nonumber
\eea
Using Eq.~(\ref{eq:AxionicFlux-standardflux}) for our toroidal construction, one finds that there are 128 axionic fluxes corresponding to 128 standard fluxes, and one can solve this set of linear relations to determine one set of fluxes from the other,
\bea
\label{eq:usualFlux2AxionFlux}
& & \left\{F, H, Q, P, P', Q', H', F' \right\} \Longleftrightarrow  \left\{\mathbb F, \mathbb H, \mathbb Q, \mathbb P, \mathbb P', \mathbb Q', \mathbb H', \mathbb F' \right\}.
\eea 
Subsequently using the ${N} = 1$ formula (\ref{eq:Vtot}) for this toroidal model, one can compute the scalar potential  induced by the generalized flux superpotential (\ref{eq:W-all-flux}) having 128 terms. The final result consists of 76276 terms in the scalar potential, in a so-called bilinear formulation presented in \cite{Carta:2016ynn,Herraez:2018vae,Escobar:2018rna,Marchesano:2019hfb,Marchesano:2020uqz,Marchesano:2021gyv}. All these 76276 terms can be distributed into 36 types of pieces, in the form of diagonal and off-diagonal flux bilinears arising from $\left\{F, H, Q, P, P', Q', H', F' \right\}$. As an immediate illustration of the significant use of the axionic fluxes (\ref{eq:AxionicFlux-standardflux}), let us mention that the scalar potential can be equivalently expressed by 10888 terms if we use the axionic flux components, distributed into the new 36 pieces arising as bilinears from $\left\{\mathbb F, \mathbb H, \mathbb Q, \mathbb P, \mathbb P', \mathbb Q', \mathbb H', \mathbb F' \right\}$. 

It is not very illuminating to display all the 36 pieces of the scalar potential which are detailed in \cite{Leontaris:2023lfc}, however to appreciate the importance of the axionic-flux polynomials we present the taxonomy of various pieces from the perspective of counting terms in scalar potential written with usual fluxes or the axionic fluxes in Table \ref{tab_term-counting}.
\begin{table}[h!]
\begin{center}
\begin{tabular}{|c||c|c||c|c|} 
\hline
%& &&&\\
& Fluxes & $\#$(terms) in $V$ using & Axionic-fluxes & $\#$(terms) in $V$ using \\
& & standard fluxes & & axionic-fluxes in (\ref{eq:AxionicFlux}) \\
\hline
%& &&&\\
1 & $F$  & 76 & ${\mathbb F}$ & 76  \\
%& & & & \\
%& & & & \\
\hline
2 & $F, H$ & 361 & ${\mathbb F}, {\mathbb H}$ & 160  \\
%& & & & \\
%& & & & \\
\hline
3 & $F, H, Q$ & 2422 & ${\mathbb F}, {\mathbb H}, {\mathbb Q}$ & 772  \\
%& &&&\\
%& &&&\\
\hline
4 & $F, H, Q, P$ & 9661 & ${\mathbb F}, {\mathbb H}, {\mathbb Q}, {\mathbb P}$ & 2356  \\
%& &&&\\
%& &&&\\
\hline
5 & $F, H, Q, P, P'$  & 23314 & ${\mathbb F}, {\mathbb H}, {\mathbb Q}, {\mathbb P}, {\mathbb P}'$ & 4855  \\
%& &&&\\
%& &&&\\
\hline
6 & $F, H, Q, P$  & 50185 & ${\mathbb F}, {\mathbb H}, {\mathbb Q}, {\mathbb P},$ & 8326  \\
& $P', Q'$ & &  ${\mathbb P}', {\mathbb Q}'$ & \\
%& &&&\\
\hline
7 & $F, H, Q, P,$  & 60750 & ${\mathbb F}, {\mathbb H}, {\mathbb Q}, {\mathbb P},$ & 9603  \\
& $P', Q', H'$ & &  ${\mathbb P}', {\mathbb Q}', {\mathbb H}'$ &\\
\hline
%& & & & \\
8 & $F, H, Q, P, $  & 76276 & ${\mathbb F}, {\mathbb H}, {\mathbb Q}, {\mathbb P},$ & 10888  \\
& $P', Q', H', F'$ & & ${\mathbb P}', {\mathbb Q}', {\mathbb H}', {\mathbb F}'$ &\\
\hline
\end{tabular}
\end{center}
\caption{Counting of scalar potential terms for a set of fluxes being turned-on at a time. \cite{Leontaris:2023lfc, Leontaris:2023mmm}}
\label{tab_term-counting}
\end{table}

\noindent
Another point worth mentioning here is the fact that all the RR axionic dependencies are encoded in the axionic fluxes (\ref{eq:AxionicFlux-standardflux}) and the scalar potential does not have an explicit dependence on any of the RR axions. In our detailed analysis of rewriting the scalar potential pieces we find that using the set of axionic fluxes reduces the number of scalar potential terms quite significantly. This subsequently helps us in understanding the insights within each terms towards seeking a compact and concise formulation of the full scalar potential and we can take an educated route to arrive at generic setups beyond the torodial orientifold. This is what is the motivation for the symplectic formulation we present next.

%%%%%%%%%%%%%%%%%%%%%%%%%%%%%%%%%%%%%%%%%%%%%%%%%%%%%%%%%%%%%%%%%%%%%%%%%%%%%%%%%%%%%%%%%%
%%%%%%%%%%%%%%%%%%%%%%%%%%%%%%%%%%%%%%%%%%%%%%%%%%%%%%%%%%%%%%%%%%%%%%%%%%%%%%%%%%%%%%%%%%

\section{Symplectic formulation}
\label{sec_symplectic}

\noindent
To begin with, let us also recollect some relevant ingredients for rewriting the F-term scalar potential in a symplectic formulation. We start by resetting our notations and conventions for symplectic formulation which would be valid for setups beyond the toroidal orientifolds.

\subsection{Necessary cohomological and symplectic ingredients}

\noindent
The massless states in the four dimensional (4D) effective theory are in one-to-one correspondence with harmonic forms which are either  even or odd under the action of an isometric, holomorphic involution ($\sigma$) acting on the internal compactifying sixfold ${X}$, and these do generate the equivariant  cohomology groups $H^{p,q}_\pm (X)$. For that purpose, let us fix our conventions, and denote the bases  of even/odd two-forms as $(\mu_\alpha, \, \nu_a)$ and four-forms as $(\tilde{\mu}_\alpha, \, \tilde{\nu}_a)$ where $\alpha\in h^{1,1}_+(X), \, a\in h^{1,1}_-(X)$. However for our current objectives, we will be interested in orientifold setups resulting in $h^{1,1}_-(X) = 0$ which means that the so-called odd-moduli $G^a$ counted by $h^{1,1}_-(X) \neq 0$ will be absent, and explicit construction of such CY orientifolds with odd two-cycles can be found in \cite{Lust:2006zg,Lust:2006zh,Blumenhagen:2008zz,Cicoli:2012vw,Gao:2013rra,Gao:2013pra}. In addition, we consider the orientifold involutions resulting in O3/O7 system with a trivial $H^3_+(X)$, and basis of 3-forms in the odd (2,1)-cohomology sector being denoted as $({\cal A}_\Lambda, {\cal B}^\Delta)$ where $\{\Lambda, \Delta\} \in \{0,1,..,h^{2,1}_-(X)\}$. We fix the normalization in the various cohomology bases as,
\bea
\label{eq:intersection}
& & \int_X \, \mu_\alpha \wedge \tilde{\mu}^\beta = {\delta}_\alpha^{\, \, \, \beta} , \quad \int_X \, \mu_\alpha \wedge \mu_\beta \wedge \mu_\gamma = \kappa_{\alpha \beta \gamma}, \quad \int_X {\cal A}_\Lambda \wedge {\cal B}^\Delta = \delta_\Lambda{}^\Delta.
\eea
Subsequently the K\"{a}hler form $J$, and the RR four-form $C_4$ can be expanded as \cite{Grimm:2004uq}: $J = t^\alpha\, \mu_\alpha,\, C_4 \simeq {\rho}_{\alpha} \, \tilde\mu^\alpha$ where $t^\alpha$, and $\rho_\alpha$ denote the Einstein-frame two-cycle volume moduli, and axions descending from the 4-form potential $C_4$. For the choice of involution $\sigma$ we have $\sigma^*\Omega_3 = - \Omega_3$, where $\Omega_3$ denotes the nowhere vanishing holomorphic three-form depending on the complex structure moduli $U^{i}$ counted in the $h^{2,1}_-(X)$ cohomology, which can be generically given as below, 
\bea
\label{eq:Omega3}
& &  \Omega_3\, \equiv  {\cal X}^\Lambda \, {\cal A}_\Lambda - \, {\cal F}_{\Lambda} \, {\cal B}^\Lambda \,. 
\eea
Here, the period vectors $({\cal X}^\Lambda, {\cal F}_\Lambda)$ are encoded in a pre-potential (${\cal F}$) of the following form,
\bea
\label{eq:prepotential}
& & \hskip-1cm {\cal F} = ({\cal X}^0)^2 \, \, f({U^i}) \,, \qquad  f({U^i}) = - \frac{1}{6}\,{{\ell}_{ijk} \, U^i\, U^j \, U^k} +  \frac{1}{2} \,{a_{ij} \, U^i\, U^j} +  \,b_{i} \, U^i +  \frac{i}{2}\, \gamma.
\eea 
In fact, the function $f({U^i})$ can generically have an infinite series of non-perturbative contributions, which we ignore for the current work assuming to be working in the large complex structure limit. The quantities $a_{ij}, b_i$ and $\gamma$ are real numbers where $\gamma$ is related to the perturbative $(\alpha^\prime)^3$-corrections on the mirror side \cite{Hosono:1994av,Arends:2014qca, Blumenhagen:2014nba}. Furthermore, the chiral coordinates $U^i$'s are defined as $U^i =\frac{\delta^i_\Lambda \, {\cal X}^\Lambda}{{\cal X}^0}$ where ${\ell}_{ijk}$'s are triple intersection numbers on the mirror (CY) threefold.

In addition to the normalizations defined in (\ref{eq:intersection}) we define the following Hodge star operations acting on the various (odd) 3-forms via introducing a set of so-called ${\cal M}$ matrices \cite{Ceresole:1995ca}, 
\begin{eqnarray}
\label{stardef}
&& \star \, {\cal A}_\Lambda =  {\cal M}_{\Lambda}^{\, \, \, \, \, \Sigma} \, \, {\cal A}_\Sigma + {\cal M}_{\Lambda \Sigma}\, \, {\cal B}^\Sigma, \, \, \, {\rm and} \, \, \, \, \star\, {\cal B}^\Lambda = {\cal M}^{\Lambda \Sigma} \, \, {\cal A}_\Sigma + {\cal M}^\Lambda_{\, \, \, \, \Sigma} \, \, {\cal B}^\Sigma\,,
\end{eqnarray}
where
\begin{eqnarray}
\label{coff}
&& {\cal M}^{\Lambda \Delta} = {\rm Im{\cal N}}^{\Lambda \Delta}, \qquad \qquad \qquad {\cal M}_{\Lambda}^{\, \, \, \, \, \Delta}  = {\rm Re{\cal N}}_{\Lambda \Gamma} \, \, {\rm Im{\cal N}}^{\Gamma \Delta}, \\
&& {\cal M}^\Lambda_{\, \, \, \, \Delta} =- \left({\cal M}_{\Lambda}^{\, \, \, \, \, \Delta}\right)^{T} , \qquad \qquad {\cal M}_{\Lambda \Delta}\, =  -{\rm Im{\cal N}}_{\Lambda \Delta} -{\rm Re{\cal N}}_{\Lambda \Sigma} \, \, {\rm Im{\cal N}}^{\Sigma \Gamma}\, \, {\rm Re{\cal N}}_{\Gamma \Delta}. \nonumber
\end{eqnarray}
Here the period matrix ${\cal N}$ for the involutively odd (2,1)-cohomology sector can be given by the derivatives of the prepotential (\ref{eq:prepotential}) as below,
\bea
\label{eq:periodN}
& & {\cal N}_{\Lambda\Delta} = \ov{\cal F}_{\Lambda\Delta} + 2 \, i \, \frac{Im({\cal F}_{\Lambda\Gamma}) \, {\cal X}^\Gamma X^\Sigma \, (Im{\cal F}_{\Sigma \Delta}) }{Im({\cal F}_{\Gamma\Sigma}) {\cal X}^\Gamma X^\Sigma}.
\eea
Moreover, it was observed in \cite{Shukla:2015hpa} that an interesting relation analogous to the period matrix expression (\ref{eq:periodN}) holds which is given as below:
\bea
\label{eq:periodF}
& & {\cal F}_{\Lambda\Delta} = \ov{\cal N}_{\Lambda\Delta} + 2 \, i \, \frac{Im({\cal N}_{\Lambda\Gamma}) \, {\cal X}^\Gamma X^\Sigma \, (Im{\cal N}_{\Sigma \Delta}) }{Im({\cal N}_{\Gamma\Sigma}) {\cal X}^\Gamma X^\Sigma},
\eea
and, similar to the period matrices given in (\ref{coff}), one can define another set of symplectic quantities as below,
\begin{eqnarray}
\label{coff2}
&& {\cal L}^{\Lambda \Delta} = {\rm Im{\cal F}}^{\Lambda \Delta},\qquad \qquad \qquad {\cal L}_{\Lambda}^{\, \, \, \, \, \Delta}  = {\rm Re{\cal F}}_{\Lambda \Gamma} \, \, {\rm Im{\cal F}}^{\Gamma \Delta}, \\
&& {\cal L}^\Lambda_{\, \, \, \, \Delta} =- \left({\cal L}_{\Lambda}^{\, \, \, \, \, \Delta}\right)^{T}, \qquad \qquad  {\cal L}_{\Lambda \Delta}\, =  -{\rm Im{\cal F}}_{\Lambda \Delta} -{\rm Re{\cal F}}_{\Lambda \Sigma} \, \, {\rm Im{\cal F}}^{\Sigma \Gamma}\, \, {\rm Re{\cal F}}_{\Gamma \Delta}. \nonumber
\end{eqnarray}
Let us mention here at the outset that in addition to having $h^{2,1}_+(X) = 0 = h^{1,1}_-(X)$ in order to avoid the D-term effects and the presence of odd moduli ($G^a$), in our current analysis we will make another assumption in our superpotential by considering the fluxes to adopt appropriate rational shifts \cite{Blumenhagen:2014nba,Escobar:2018rna} in order to absorb the appearance of $a_{ij}$ and $b_i$ sourced from the prepotential (\ref{eq:prepotential}). Also, we set $\gamma = 0$ assuming the large complex structure limit.

%%%%%%%%%%%%%%%%%%%%%%%%%%%%%%%%%%%%%%%%%%%%%%%%%%%%%%%%%%%%%%%%%%%%%%%%%%%%%%%%%%%%%%%%%%
%%%%%%%%%%%%%%%%%%%%%%%%%%%%%%%%%%%%%%%%%%%%%%%%%%%%%%%%%%%%%%%%%%%%%%%%%%%%%%%%%%%%%%%%%%

\subsection{The K\"ahler potential and the flux superpotential}

\noindent
Using these pieces of information one defines a set of chiral coordinates, namely $S, T_\alpha$ and $U^i$, as below \cite{Benmachiche:2006df},
\bea
\label{eq:N=1_coords}
& & \hskip-2cm S \equiv C_0 + \, i \, e^{-\phi} = C_0 + i\, s, \quad T_\alpha= \rho_\alpha - i\, \tau_\alpha, \quad U^i = v^i - i\, u^i,
\eea
where $\{C_0, \rho_\alpha, v^i\}$ are axionic moduli while $\{s, \tau_\alpha, u^i\}$ are saxionic moduli. The Einstein-frame overall volume (${\cal V}$) of the internal background is connected with the 2-cycle volume moduli $t^\alpha$ and the 4-cycle volume moduli as below,
\begin{eqnarray}
& & {\cal V} = \frac{1}{6} \kappa_{\alpha \beta \gamma} \, t^\alpha\, t^\beta \, t^{\gamma}, \qquad \tau_\alpha = \partial_{t^\alpha} {\cal V} = \frac{1}{2} \kappa_{\alpha\beta\gamma} t^\beta t^\gamma.
\end{eqnarray}
Using appropriate chiral variables ($S, T_\alpha, U^i$) as defined in (\ref{eq:N=1_coords}), a generic form of the tree-level K\"{a}hler potential can be written as below,
\bea
\label{eq:K}
& & \hskip-1.5cm K = -\ln\left(-\,i\int_{X}\Omega_3\wedge{\bar\Omega_3}\right) - \ln\left(-i(S-\ov S)\right) -2\ln{\cal V}.%\\
%& & \hskip-1cm = -\ln \left(\frac{4}{3} \, \ell_{ijk}\, u^i u^j u^k + 2\, \gamma \right)\, - \, \ln (2\,s) - 2 \ln\left(\frac{1}{6} \,{\ell_{\alpha \beta \gamma} \, t^\alpha\, t^\beta \, t^{\gamma}}\right). \nonumber
\eea
The tree-level K\"ahler potential (\ref{eq:K}) leads to a decoupled sector for $U^i$ moduli  and the $\{S, T_\alpha\}$ moduli sectors, thus  implying the following derivatives along with the block diagonals matrices,
\bea
\label{eq:derK-Kalphabeta}
& & K_S = \frac{i}{2\,s} = - K_{\ov{S}}, \quad K_{T_\alpha} = -\frac{i\, t^\alpha}{2{\cal V}} = -K_{{\ov T}_\alpha}, \\
& & K_{S\ov{S}} = \frac{1}{4\, s^2}, \quad K_{S\ov{T}_\alpha} = 0 = K_{T_\alpha\ov{S}}, \quad K_{T_\alpha{\ov T}_\beta} = 4\left(\tau_\alpha\tau_\beta - {\cal V} \kappa_{\alpha\beta}\right) \equiv 4\, {\cal G}_{\alpha\beta},\nonumber\\
& & K^{S\ov{S}} = 4\, s^2, \quad K^{S\ov{T}_\alpha} = 0 = K^{T_\alpha\ov{S}}, \quad K^{T_\alpha{\ov T}_\beta} = \frac{1}{16{\cal V}^2}\left(2\, t^\alpha t^\beta-4{\cal V}\kappa^{\alpha\beta} \right) \equiv \frac{1}{4} {\cal G}^{\alpha\beta}, \nonumber
\eea
where we have introduced two new quantities ${\cal G}_{\alpha\beta}$ and ${\cal G}^{\alpha\beta}$ to avoid any confusion with lower/upper indices in the K\"ahler metric and its inverse metric. We also note that the metrics defined in (\ref{eq:derK-Kalphabeta}) result in the following useful identities \cite{Biswas:2024ewk},
\bea
\label{eq:metric-identities}
& & \hskip-1cm {\cal G}_{\alpha \beta} t^\alpha = {\cal V}\,\tau_\beta, \quad {\cal G}_{\alpha \beta} t^\alpha t^\beta = 3 {\cal V}^2, \quad {\cal G}^{\alpha \beta} \tau_\alpha = \frac{t^\alpha}{\cal V}, \quad {\cal G}^{\alpha \beta} \tau_\alpha \tau_\beta = 3.
\eea
Now, it turns out that making successive applications of T/S-dualities results in the need of introducing more and more fluxes. In fact, one needs a total of four S-dual pairs of fluxes, commonly denotes as:  $(F, H), \, (Q, P), \, (P^\prime, Q^\prime)$ and $(H^\prime, F^\prime)$ \cite{Aldazabal:2006up,Aldazabal:2008zza,Shukla:2015rua,Aldazabal:2006up,Aldazabal:2008zza,Aldazabal:2010ef,Lombardo:2016swq,Lombardo:2017yme,Leontaris:2023lfc}. This leads to the following generalized flux superpotential,
\bea
\label{eq:W-all-4}
& & \hskip-1cm W = \int_{X} \biggl[\left(F - S\, H \right) + \left(Q - S\, P \right)\triangleright {\cal J} + \left(P^\prime - S\, Q^\prime \right)\diamond {\cal J}^2 + \left(H^\prime - S\, F^\prime \right)\odot {\cal J}^3\biggr] \wedge \Omega_3\,,
\eea
where ${\cal J} = T_\alpha\, \mu^\alpha$ the dual complexified four-form dual to the K\"ahler form, and the symplectic form of the various flux actions are defined as below \cite{Leontaris:2023mmm},
\bea
\label{eq:FluxActions-3}
&& \left(Q\triangleright {\cal J} \right) = T_\alpha \, {Q}^{\alpha}, \qquad \qquad \qquad \qquad \, \, \left(P\triangleright {\cal J} \right) = T_\alpha \, {P}^{\alpha} \, ,\\
&& \left(P^\prime \diamond {\cal J}^2 \right)  = \frac{1}{2} \, {P}^{\prime\beta\gamma}\, T_\beta \, T_\gamma \,, \qquad  \qquad \quad \left(Q^\prime \diamond {\cal J}^2 \right)  = \frac{1}{2} \, {Q}^{\prime\beta\gamma}\, T_\beta \, T_\gamma\,, \nonumber\\
&& \left(H^\prime \odot {\cal J}^3 \right) = \frac{1}{3 !} \, {H}^{\prime\alpha\beta\gamma}\,T_\alpha \, T_\beta \, T_\gamma\, , \qquad \, \, \, \left(F^\prime \odot {\cal J}^3 \right) = \frac{1}{3 !} \, {F}^{\prime\alpha\beta\gamma}\,T_\alpha \, T_\beta \, T_\gamma. \nonumber
\eea
Here ${Q}^{\alpha}, {P}^{\alpha}, {P}^{\prime\beta\gamma}, {Q}^{\prime\beta\gamma}, {H}^{\prime\alpha\beta\gamma}$ and ${F}^{\prime\alpha\beta\gamma}$ denote the 3-forms similar to the standard $F_3/H_3$ fluxes and can be expanded in the symplectic basis $\{{\cal A}_\Lambda, {\cal B}^\Delta\}$ in the following manner,
\bea
\label{eq:FluxActions-2}
&& F = F^\Lambda \, {\cal A}_\Lambda - F_{\Lambda} \, {\cal B}^\Lambda\,,\qquad \qquad \qquad \qquad H = H_\Lambda \, {\cal A}_\Lambda - H_{\Lambda} \, {\cal B}^\Lambda\,,\\
&& {Q}^{\alpha} = {Q}^{\alpha{}\Lambda} \, {\cal A}_\Lambda - {Q}^{\alpha}{}_{\Lambda} \, {\cal B}^\Lambda\,,\qquad \qquad \qquad {P}^{\alpha} = {P}^{\alpha{}\Lambda} \, {\cal A}_\Lambda - {P}^{\alpha}{}_{\Lambda} \, {\cal B}^\Lambda\,,\nonumber\\
&& {P}^{\prime\beta\gamma}  = {P}^{\prime\beta\gamma{}\Lambda} \, {\cal A}_\Lambda - {P}^{\prime\beta\gamma}{}_{\Lambda} \, {\cal B}^\Lambda\,, \qquad \qquad {Q}^{\prime\beta\gamma}\,= {Q}^{\prime\beta\gamma{}\Lambda} \, {\cal A}_\Lambda - {Q}^{\prime\beta\gamma}{}_{\Lambda} \, {\cal B}^\Lambda\,, \nonumber\\
&& H^{\prime\alpha\beta\gamma} =  {H}^{\prime\alpha\beta\gamma{}\Lambda} \, {\cal A}_\Lambda - {H}^{\prime\alpha\beta\gamma}{}_{\Lambda} \, {\cal B}^\Lambda\, , \qquad F^{\prime\alpha\beta\gamma}\, = {F}^{\prime\alpha\beta\gamma{}\Lambda} \, {\cal A}_\Lambda - {F}^{\prime\alpha\beta\gamma}{}_{\Lambda} \, {\cal B}^\Lambda\,. \nonumber
\eea
Note that the flux components $Q^\alpha$ and $P^\alpha$ having a single index $\alpha \in h^{1,1}_+$ (and $\Lambda \in h^{2,1}_-$ indices) are counted accordingly. However, counting of the prime flux components is a bit tricky because their appearance as ${P}^{\prime\beta\gamma}, {Q}^{\prime\beta\gamma}, {H}^{\prime\alpha\beta\gamma}$ and ${F}^{\prime\alpha\beta\gamma}$ in (\ref{eq:FluxActions-3}) does not seem to have any constraint except for being symmetric in indices $\{\alpha, \beta, \gamma\} \in h^{1,1}_+$. For that purpose, motivated by the previous toroidal studies \cite{Aldazabal:2010ef, Lombardo:2016swq, Lombardo:2017yme, Leontaris:2023lfc}, one can equivalently define \cite{Leontaris:2023mmm},
\bea
\label{eq:primefluxIIBa}
& & P^\prime_{\alpha\Lambda}, \, {P^\prime}_\alpha{}^\Lambda, \qquad Q^\prime_{\alpha\Lambda}, \, {Q^\prime}_\alpha{}^\Lambda, \qquad H^\prime_{\Lambda}, \, {H^\prime}^\Lambda, \qquad F^\prime_{\Lambda}, \, {F^\prime}^\Lambda\,,
\eea
which are related to the earlier set of prime fluxes as below
\bea
\label{eq:primefluxIIBc}
& & {P}^{\prime\beta\gamma{}\Lambda}= {P^\prime}_{\alpha}{}^{\Lambda}\, \kappa^{\alpha\beta\gamma}, \qquad \, {P}^{\prime\beta\gamma}{}_{\Lambda} = {P^\prime}_{\alpha \Lambda} \, \kappa^{\alpha\beta\gamma}\,,\\
& & {Q}^{\prime\beta\gamma{}\Lambda}= {Q^\prime}_{\alpha}{}^{\Lambda}\, \kappa^{\alpha\beta\gamma}, \qquad \, {Q}^{\prime\beta\gamma}{}_{\Lambda} = {Q^\prime}_{\alpha \Lambda} \, \kappa^{\alpha\beta\gamma}\,,\nonumber\\
& & {H}^{\prime\alpha\beta\gamma{}\Lambda}= {H^\prime}^{\Lambda}\, \kappa^{\alpha\beta\gamma}, \qquad {H}^{\prime\alpha\beta\gamma}{}_{\Lambda} = {H^\prime}_{\Lambda} \, \kappa^{\alpha\beta\gamma}\,, \nonumber\\
& & {F}^{\prime\alpha\beta\gamma{}\Lambda}= {F^\prime}^{\Lambda}\, \kappa^{\alpha\beta\gamma}, \qquad \, \, {F}^{\prime\alpha\beta\gamma}{}_{\Lambda} = {F^\prime}_{\Lambda} \, \kappa^{\alpha\beta\gamma}.\nonumber
\eea
Here $\kappa^{\alpha\beta\gamma}$ is defined as $\kappa^{\alpha\beta\gamma} = {\cal V}^3\, \kappa_{\alpha^\prime\beta^\prime\gamma^\prime}\,{\cal G}^{\alpha\alpha^\prime}\, {\cal G}^{\beta\beta^\prime} \, {\cal G}^{\gamma\gamma^\prime}$ where the tree-level metric ${\cal G}^{\alpha\beta}$ and its inverse ${\cal G}_{\alpha\beta}$ are introduced in (\ref{eq:derK-Kalphabeta}). This leads to the following interesting relation \cite{Leontaris:2023mmm}
\bea
\label{eq:metric-identities}
& & \hskip-1.0cm t^\alpha = \frac{1}{2} \, \kappa^{\alpha\beta\gamma}\, \tau_\beta \, \tau_\gamma, \quad {\cal V} = \frac{1}{3!}\, \kappa^{\alpha\beta\gamma} \, \tau_\alpha \, \tau_\beta\, \tau_\gamma,\nonumber
\eea
where $\tau_\alpha$ corresponds to the volume of the 4-cycle and can be written in terms of the 2-cycle volumes as: $\tau_\alpha = \frac{1}{2}\, \kappa_{\alpha\beta\gamma}\, t^\beta\, t^\gamma = \frac{1}{2}\, \kappa_\alpha$. Here we used the shorthand notations $\kappa_{\alpha} = \kappa_{\alpha\beta}\,t^\beta = \kappa_{\alpha\beta\gamma}\,t^\beta\, t^\gamma$ etc.

From this approach it is clear that counting of prime fluxes is also similar to those of the non-prime fluxes, i.e. $\{H', F'\}$ are counted by the $h^{2,1}_-({\rm CY})$ indices similar to the standard $H_3/F_3$ fluxes while $\{P', Q'\}$ are counted via one $\alpha$ index and one $\Lambda$ index. However, the crucial property we must take into account is  the holomorphicity of the flux superpotential which shows that fluxes used in actions (\ref{eq:FluxActions-3}) and (\ref{eq:FluxActions-2}) are the ones to be considered in the superpotential along with the chiral variables $S, T_\alpha$ and $U^i$. Subsequently one has the following generalized flux superpotential,
\bea
\label{eq:W-all-1}
& & \hskip-1.0cm W  = \int_{X} \Biggl[\left({F}  + {Q}^{\alpha}\, T_\alpha + \frac{1}{2} {P}^{\prime\alpha\beta}\, T_\alpha\, T_\beta \, +  \, \frac{1}{6} H^{\prime\alpha\beta\gamma}\, T_\alpha \, T_\beta \, T_\gamma \right) \\
& & \hskip1cm - \, S \, \left({H}  + {P}^{\alpha}\, T_\alpha +  \frac{1}{2} {Q}^{\alpha\beta}\, T_\alpha\, T_\beta \, +\, \frac{1}{6} \,F^{\prime\alpha\beta\gamma}\, T_\alpha \, T_\beta \, T_\gamma \right) \Biggr]_3 \wedge \Omega_3 \,,\nonumber
\eea
where all the terms appearing inside the bracket $[...]_3$ denote a collection of three-forms as expanded in (\ref{eq:FluxActions-2}). Let us also make a side remark that for the holomorphicity of the flux superpotential all these 3-form flux ingredients in (\ref{eq:W-all-1}), or equivalently in (\ref{eq:FluxActions-2}), are independent of the volume moduli and therefore such fluxes are used for expressing the superpotential for the purpose of computing the scalar potential using the $N = 1$ formula (\ref{eq:Vtot}). After the scalar potential is computed one can always reshuffle and rewrite the scalar potential pieces in order to achieve a more compact way of expressing the full potential, which a priori is a complicated task.

Using the flux actions in Eq.~(\ref{eq:FluxActions-2}), the generalized flux superpotential $W$ in (\ref{eq:W-all-1}) can be equivalently given as,
\bea
\label{eq:W-gen}
& & \hskip-1.0cm W \equiv \int_{X} [{\rm Flux}]_3 \wedge \Omega_3 = e_\Lambda \, {\cal X}^\Lambda - m^\Lambda \, {\cal F}_\Lambda\,,
\eea
where the symplectic electromagnetic vectors $(e_\Lambda, m^\Lambda)$ are given as,
\begin{eqnarray}
\label{eq:eANDm}
& &  e_\Lambda = \left({F}_\Lambda - S \, {H}_\Lambda\right)  + T_\alpha \left({Q}^{\alpha}{}_\Lambda\,  - S \, {P}^{\alpha}{}_\Lambda \right) + \frac{1}{2} T_\alpha\, T_\beta \left({P}^{\prime\alpha\beta}{}_\Lambda\, - S\, {Q}^{\prime\alpha\beta}{}_\Lambda \right) \, \\
& & \hskip2cm + \, \frac{1}{6} \, T_\alpha \, T_\beta \, T_\gamma \left(H^{\prime\alpha\beta\gamma}{}_\Lambda\, - S \, F^{\prime\alpha\beta\gamma}{}_\Lambda\right), \, \nonumber\\
& &  m^\Lambda = \left({F}^\Lambda - S \, {H}^\Lambda\right)  + T_\alpha \left({Q}^{\alpha}{}^\Lambda\,  - S \, {P}^{\alpha}{}^\Lambda \right) + \frac{1}{2} T_\alpha\, T_\beta \left({P}^{\prime\alpha\beta}{}^\Lambda\, - S\, {Q}^{\prime\alpha\beta}{}^\Lambda \right) \, \nonumber\\
& & \hskip2cm + \, \frac{1}{6} \, T_\alpha \, T_\beta \, T_\gamma \left(H^{\prime\alpha\beta\gamma}{}^\Lambda\, - S \, F^{\prime\alpha\beta\gamma}{}^\Lambda\right).\nonumber
\end{eqnarray}
The generic superpotential (\ref{eq:W-gen}) is linear in the axio-dilaton modulus $S$ and has cubic dependence in moduli $U^i$ and $T_\alpha$ both. Using the symplectic vectors $(e_\Lambda, m^\Lambda)$ of (\ref{eq:eANDm}) in the flux superpotential (\ref{eq:W-gen}), one can compute the derivatives with respect to chiral variables, $S$ and $T_\alpha$ which are given by
\begin{eqnarray}
\label{eq:derW_gen}
& & W_S = {(e_1)}_\Lambda \, {\cal X}^\Lambda - {(m_1)}^\Lambda \, {\cal F}_\Lambda, \qquad W_{T_\alpha} = {(e_2)}^\alpha_\Lambda \, {\cal X}^\Lambda - {(m_2)}^\alpha{}^\Lambda \, {\cal F}_\Lambda,
\end{eqnarray}
where the two new pairs of symplectic vectors $(e_1, m_1)$ and $(e_2, m_2)$ are given as:
\begin{eqnarray}
\label{eq:e1ANDm1}
& &  {(e_1)}_\Lambda = - \biggl[{H}_\Lambda  + T_\alpha \, {P}^{\alpha}{}_\Lambda + \frac{1}{2} T_\alpha\, T_\beta \, {Q}^{\prime\alpha\beta}{}_\Lambda + \, \frac{1}{6} \, T_\alpha \, T_\beta \, T_\gamma \, F^{\prime\alpha\beta\gamma}{}_\Lambda \biggr], \,\\
& &  {(m_1)}^\Lambda = -\biggl[{H}^\Lambda + T_\alpha \, {P}^{\alpha}{}^\Lambda + \frac{1}{2} T_\alpha\, T_\beta \, {Q}^{\prime\alpha\beta}{}^\Lambda  + \, \frac{1}{6} \, T_\alpha \, T_\beta \, T_\gamma  \, F^{\prime\alpha\beta\gamma}{}^\Lambda\, \biggr]\,,\nonumber
\end{eqnarray}
and
\begin{eqnarray}
\label{eq:e2ANDm2}
& & \hskip-1.5cm {(e_2)}^\alpha_\Lambda = \left({Q}^{\alpha}{}_\Lambda - S \, {P}^{\alpha}{}_\Lambda \right) + T_\beta \left({P}^{\prime\alpha\beta}{}_\Lambda\, - S\, {Q}^{\prime\alpha\beta}{}_\Lambda \right) + \, \frac{1}{2} T_\beta \, T_\gamma \left(H^{\prime\alpha\beta\gamma}{}_\Lambda\, - S \, F^{\prime\alpha\beta\gamma}{}_\Lambda\right),\\
& & \hskip-1.5cm {(m_2)}^\alpha{}^\Lambda = \left({Q}^{\alpha}{}^\Lambda\,  - S \, {P}^{\alpha}{}^\Lambda \right) + \, T_\beta \left({P}^{\prime\alpha\beta}{}^\Lambda\, - S\, {Q}^{\prime\alpha\beta}{}^\Lambda \right) + \, \frac{1}{2} \, T_\beta \, T_\gamma \left(H^{\prime\alpha\beta\gamma}{}^\Lambda\, - S \, F^{\prime\alpha\beta\gamma}{}^\Lambda\right).\nonumber
\end{eqnarray}

\subsection{Invoking the axionic fluxes}

\noindent
With the concrete form of the holomorphic superpotential at hand, now, we define the following set of the so-called axionic-flux combinations which will turn out to be extremely useful for rearranging the scalar potential pieces into a compact form,
\bea
\label{eq:AxionicFlux}
& & \hskip-1cm {\mathbb F}_\Lambda = {F}_\Lambda  + \rho_\alpha \, {Q}^{\alpha}{}_\Lambda + \frac{1}{2} \rho_\alpha\, \rho_\beta \, {P}^{\prime\alpha\beta}{}_\Lambda + \, \frac{1}{6} \, \rho_\alpha \, \rho_\beta \, \rho_\gamma \, H^{\prime\alpha\beta\gamma}{}_\Lambda - \, C_0 \, {\mathbb H}_\Lambda\,, \\
& & \hskip-1cm {\mathbb H}_\Lambda = {H}_\Lambda  + \rho_\alpha \, {P}^{\alpha}{}_\Lambda + \frac{1}{2} \rho_\alpha\, \rho_\beta \, {Q}^{\prime\alpha\beta}{}_\Lambda + \, \frac{1}{6} \, \rho_\alpha \, \rho_\beta \, \rho_\gamma \, F^{\prime\alpha\beta\gamma}{}_\Lambda,\nonumber\\
& & \hskip-1cm {{\mathbb Q}}^{\alpha}{}_\Lambda = \, {Q}^{\alpha}{}_\Lambda + \, \rho_\beta \, {P}^{\prime\alpha\beta}{}_\Lambda + \, \frac{1}{2} \, \rho_\beta \, \rho_\gamma \, H^{\prime\alpha\beta\gamma}{}_\Lambda -\, C_0\, { {\mathbb P}}^{\alpha}{}_\Lambda, \nonumber\\ 
& & \hskip-1cm {{\mathbb P}}^{\alpha}{}_\Lambda = \, {P}^{\alpha}{}_\Lambda + \, \rho_\beta \, {Q}^{\prime\alpha\beta}{}_\Lambda + \, \frac{1}{2} \, \rho_\beta \, \rho_\gamma \, F^{\prime\alpha\beta\gamma}{}_\Lambda\,, \nonumber\\
& & \hskip-1cm {{\mathbb P}}^{\prime\alpha\beta}{}_\Lambda =  {P}^{\prime\alpha\beta}{}_\Lambda + \, \rho_\gamma \, H^{\prime\alpha\beta\gamma}{}_\Lambda\, - C_0\, { {\mathbb Q}}^{\prime\alpha\beta}{}_\Lambda\,, \nonumber\\
& & \hskip-1cm {{\mathbb Q}}^{\prime\alpha\beta}{}_\Lambda =  {Q}^{\prime\alpha\beta}{}_\Lambda + \, \rho_\gamma \, F^{\prime\alpha\beta\gamma}{}_\Lambda, \nonumber\\
& & \hskip-1cm {\mathbb H}^{\prime\alpha\beta\gamma}{}_\Lambda = \, H^{\prime\alpha\beta\gamma}{}_\Lambda\, - \, C_0\,{\mathbb F}^{\prime\alpha\beta\gamma}{}_\Lambda, \nonumber\\
& & \hskip-1cm {\mathbb F}^{\prime\alpha\beta\gamma}{}_\Lambda = \, F^{\prime\alpha\beta\gamma}{}_\Lambda\,, \nonumber
\eea
where the analogous axionic fluxes with upper $\Lambda$ index are expressed in a similar fashion. Using these axionic-flux combinations (\ref{eq:AxionicFlux}) along with the definitions of chiral variables in Eq.~(\ref{eq:N=1_coords}), the three pairs of symplectic vectors, namely $(e, m), \, (e_1, m_1)$ and $(e_2, m_2)$ which are respectively given in Eqs.~(\ref{eq:eANDm}), (\ref{eq:e1ANDm1}) and (\ref{eq:e2ANDm2}), can be expressed in the following compact form,
\bea
\label{eq:em}
& & e_\Lambda = \left({\mathbb F}_\Lambda - s \, {\mathbb P}^{}{}_{\Lambda} - {\mathbb P}^{\prime}{}_{\Lambda} + s\, {\mathbb F}^{\prime}{}_{\Lambda} \right)  + i \, \left( - s \, {\mathbb H}_\Lambda - {{\mathbb Q}}^{}{}_{\Lambda} + s \, {{\mathbb Q}}^{\prime}{}_{\Lambda} + {{\mathbb H}}^{\prime}{}_{\Lambda}\right),\\
& & m^\Lambda = \left({\mathbb F}^\Lambda - s \, {\mathbb P}^{}{}^{\Lambda} - {\mathbb P}^{\prime}{}^{\Lambda} + s\, {\mathbb F}^{\prime}{}^{\Lambda} \right)  + i \, \left( -\, s \, {\mathbb H}^\Lambda - {{\mathbb Q}}^{}{}^{\Lambda} + s \, {{\mathbb Q}}^{\prime}{}^{\Lambda} + {{\mathbb H}}^{\prime}{}^{\Lambda}\right), \nonumber
\eea
\bea
\label{eq:e1m1}
& & \hskip-1.5cm {(e_1)}_\Lambda = \left(- \, {\mathbb H}_\Lambda + \, {{\mathbb Q}}^{\prime}{}_{\Lambda} \right) + i \left({\mathbb P}^{}{}_{\Lambda} - \, {\mathbb F}^{\prime}{}_{\Lambda} \right)\,, \quad {(m_1)}^\Lambda = \left(-\, {\mathbb H}^\Lambda + \, {{\mathbb Q}}^{\prime}{}^{\Lambda}\right) + i \left({\mathbb P}^{}{}^{\Lambda}  - \, {\mathbb F}^{\prime}{}^{\Lambda} \right)\,, 
\eea
\bea
\label{eq:e2m2}
& & {(e_2)}^\alpha{}_\Lambda = \left({{\mathbb Q}}^{\alpha}{}_{\Lambda} - \, s \, {{\mathbb Q}}^{\prime\alpha}{}_{\Lambda} - \, {{\mathbb H}}^{\prime\alpha}{}_{\Lambda}\right) + i \left(- s \, {\mathbb P}^{\alpha}{}_{\Lambda} - {\mathbb P}^{\prime\alpha}{}_{\Lambda} + s\, {\mathbb F}^{\prime\alpha}{}_{\Lambda} \right)\,,\\
& & {(m_2)}^\alpha{}^\Lambda = \left({{\mathbb Q}}^{\alpha}{}^{\Lambda} - s \, {{\mathbb Q}}^{\prime\alpha}{}^{\Lambda} - {{\mathbb H}}^{\prime\alpha}{}^{\Lambda}\right) + i \left( - \, s \, {\mathbb P}^{\alpha}{}^{\Lambda} - {\mathbb Q}^{\prime\alpha}{}^{\Lambda} + s\, {\mathbb F}^{\prime\alpha}{}^{\Lambda} \right)\,, \nonumber
\eea
where we have used the shorthand notations such as ${\mathbb Q}^{}_\Lambda = \tau_\alpha\,{\mathbb Q}^{\alpha}{}_\Lambda$, ${\mathbb Q}^{\prime\alpha}_\Lambda = \,\tau_\beta{\mathbb Q}^{\prime\alpha\beta}{}_\Lambda$, ${\mathbb Q}^{\prime}_\Lambda = \frac12 \tau_\alpha\,\tau_\beta{\mathbb Q}^{\prime\alpha\beta}{}_\Lambda$,  ${\mathbb H}^{\prime}_\Lambda = \frac16\,\tau_\alpha\tau_\beta \tau_\gamma{\mathbb H}^{\prime\alpha\beta\gamma}{}_\Lambda$ etc. In addition, we mention that such shorthand notations are applicable only with $\tau_\alpha$ contractions, and not to be (conf)used with axionic ($\rho_\alpha$) contractions. This convention will be used wherever the $(Q, P)$, $(P^\prime, Q^\prime)$ and $(H^\prime, F^\prime)$ fluxes are seen with/without a free index $\alpha \in h^{1,1}_+(X)$. Here we recall that $\tau_\alpha = \frac{1}{2}\, {\kappa}_{\alpha \beta \gamma} t^\beta t^\gamma$.

Now the main task is to rewrite the scalar potential (\ref{eq:Vtot}) in a generic symplectic formulation. The blocks in the K\"ahler metric and inverse metric corresponding  to the K\"ahler moduli and axio-dilaton dependent sector are determined by (\ref{eq:derK-Kalphabeta}) while for simplifying the complex structure dependent pieces one needs to use the period matrix components defined in (\ref{coff})-(\ref{coff2}). In this regard, one of the most important identities for simplifying the scalar potential is the following one \cite{Ceresole:1995ca},
\begin{eqnarray}
\label{eq:Identity1}
&& K^{i \ov j} \, (D_i {\cal X}^{\Lambda}) \, (\ov D_{\ov j} \ov{{\cal X}^{\Delta}}) = - \ov{{\cal X}^{\Lambda}} \,  {\cal X}^{\Delta} - \frac{1}{2} \, e^{-K_{cs}} \, {\rm Im{\cal N}}^{\Lambda \Delta}.
\end{eqnarray}
Subsequently after some tedious computations, one arrives at a generic scalar potential promoted for the beyond toroidal case. This scalar potential can be expressed using the following 36 pieces,
\bea
\label{eq:master1}
& & \hskip-1cm V = V_{\mathbb F \mathbb F} + V_{\mathbb H \mathbb H}+ V_{\mathbb Q \mathbb Q}+ V_{\mathbb P \mathbb P}+ V_{{\mathbb P}^\prime {\mathbb P}^\prime} + V_{{\mathbb Q}^\prime {\mathbb Q}^\prime} + V_{{\mathbb H}^\prime {\mathbb H}^\prime}+ V_{{\mathbb F}^\prime {\mathbb F}^\prime}, \\
& & \hskip-0.5cm + V_{\mathbb F \mathbb H} + V_{\mathbb F \mathbb Q} + V_{\mathbb F \mathbb P}+ V_{\mathbb F {\mathbb P}^\prime} + V_{\mathbb F {\mathbb Q}^\prime} + V_{\mathbb F {\mathbb H}^\prime} + V_{\mathbb F {\mathbb F}^\prime} + V_{\mathbb H \mathbb Q} + V_{\mathbb H \mathbb P}+ V_{\mathbb H {\mathbb P}^\prime} + V_{\mathbb H {\mathbb Q}^\prime} + V_{\mathbb H {\mathbb H}^\prime} + V_{\mathbb H {\mathbb F}^\prime}\nonumber\\
& & \hskip-0.5cm  + V_{\mathbb Q \mathbb P} + V_{\mathbb Q {\mathbb P}^\prime} + V_{\mathbb Q {\mathbb Q}^\prime}+ V_{\mathbb Q{\mathbb H}^\prime} + V_{\mathbb Q {\mathbb F}^\prime} +  V_{\mathbb P {\mathbb P}^\prime} + V_{\mathbb P {\mathbb Q}^\prime} + V_{\mathbb P {\mathbb H}^\prime} + V_{\mathbb P {\mathbb F}^\prime} + V_{{\mathbb P}^\prime {\mathbb Q}^\prime}+  V_{{\mathbb P}^\prime {\mathbb H}^\prime}+ V_{{\mathbb P}^\prime {\mathbb F}^\prime}   \nonumber\\
& & \hskip-0.5cm + V_{{\mathbb Q}^\prime {\mathbb H}^\prime} + V_{{\mathbb Q}^\prime {\mathbb F}^\prime} + V_{{\mathbb H}^\prime {\mathbb F}^\prime}\,.\nonumber
\eea
The explicit forms of the various pieces are elaborated in the appendix of \cite{Leontaris:2023mmm}.

\subsection{Master formula for the scalar potential}

\noindent
The attempts so far have just been to elaborate on the insights of various terms and how they could appear from the flux superpotential in connection with the standard U-dual flux parameters, and it is desirable that we club these 36 terms in a more concise symplectic formulation. With this goal in mind, we investigated the 36 pieces in some more detail and found that using the axionic fluxes (\ref{eq:AxionicFlux}), the following triplet $\{\chi, \psi, \Psi^\alpha\}$ of quantities turn out to be useful,
\bea
\label{eq:chi-psi-Psi}
& & \hskip-1cm \psi_\Lambda = s\, \left(- \, {\mathbb H}_\Lambda + \, {{\mathbb Q}}^{\prime}{}_{\Lambda} \right) + i \,  s \, \left({\mathbb P}^{}{}_{\Lambda} - \, {\mathbb F}^{\prime}{}_{\Lambda} \right), \\%\quad \psi^\Lambda = s\, \left(-\, {\mathbb H}^\Lambda + \, {{\mathbb Q}}^{\prime}{}^{\Lambda}\right) + i \, s \, \left({\mathbb P}^{}{}^{\Lambda}  - \, {\mathbb F}^{\prime}{}^{\Lambda} \right), \\
& & \hskip-1cm  \chi_\Lambda = \left({\mathbb F}_\Lambda - {\mathbb P}^{\prime}{}_{\Lambda} \right)  + i \, \left(- {{\mathbb Q}}^{}{}_{\Lambda} + {{\mathbb H}}^{\prime}{}_{\Lambda}\right) + i\, \psi_\Lambda, \nonumber\\%\quad \chi^\Lambda = \left({\mathbb F}^\Lambda - {\mathbb P}^{\prime}{}^{\Lambda} \right)  + i \, \left(- {{\mathbb Q}}^{}{}^{\Lambda} + {{\mathbb H}}^{\prime}{}^{\Lambda}\right) + i\,  \psi^\Lambda, \nonumber\\
& & \hskip-1cm \Psi^\alpha{}_\Lambda = \left({{\mathbb Q}}^{\alpha}{}_{\Lambda} - \, s \, {{\mathbb Q}}^{\prime\alpha}{}_{\Lambda} - \, {{\mathbb H}}^{\prime\alpha}{}_{\Lambda}\right) + i \left(- s \, {\mathbb P}^{\alpha}{}_{\Lambda} - {\mathbb P}^{\prime\alpha}{}_{\Lambda} + s\, {\mathbb F}^{\prime\alpha}{}_{\Lambda} \right),\nonumber%\\
%& & \hskip-1cm \Psi^\alpha{}^\Lambda = \left({{\mathbb Q}}^{\alpha}{}^{\Lambda} - s \, {{\mathbb Q}}^{\prime\alpha}{}^{\Lambda} - {{\mathbb H}}^{\prime\alpha}{}^{\Lambda}\right) + i \left( - \, s \, {\mathbb P}^{\alpha}{}^{\Lambda} - {\mathbb Q}^{\prime\alpha}{}^{\Lambda} + s\, {\mathbb F}^{\prime\alpha}{}^{\Lambda} \right), \nonumber
\eea
where the analogous quantities with upper $\Lambda$ index are expressed in a similar fashion.
Also we used the shorthand notations such as ${\mathbb Q}^{}_\Lambda = \tau_\alpha\,{\mathbb Q}^{\alpha}{}_\Lambda$, ${\mathbb Q}^{\prime\alpha}_\Lambda = \,\tau_\beta{\mathbb Q}^{\prime\alpha\beta}{}_\Lambda$, ${\mathbb Q}^{\prime}_\Lambda = \frac12 \tau_\alpha\,\tau_\beta{\mathbb Q}^{\prime\alpha\beta}{}_\Lambda$,  ${\mathbb H}^{\prime}_\Lambda = \frac16\,\tau_\alpha\tau_\beta \tau_\gamma{\mathbb H}^{\prime\alpha\beta\gamma}{}_\Lambda$ etc. In the similar way we write $\Psi_\Lambda = \tau_\alpha \, \Psi^\alpha{}_\Lambda$ and $\Psi^\Lambda = \tau_\alpha\, \Psi^\alpha{}^\Lambda$ wherever $\Psi$ appears without an $h^{1,1}_+$ index $\alpha$. Subsequently, we will have the following relations consistent with out shorthand notations,
\bea
\label{eq:Psi-2}
& & \Psi{}_\Lambda = \left({{\mathbb Q}}^{}{}_{\Lambda} - 2\, s \, {{\mathbb Q}}^{\prime}{}_{\Lambda} - 3\, {{\mathbb H}}^{\prime}{}_{\Lambda}\right) + i \left(- s \, {\mathbb P}^{}{}_{\Lambda} - 2\, {\mathbb P}^{\prime}{}_{\Lambda} + 3\,s\, {\mathbb F}^{\prime}{}_{\Lambda} \right)\,,\\
& & \Psi{}^\Lambda = \left({{\mathbb Q}}^{}{}^{\Lambda} - 2\,s \, {{\mathbb Q}}^{\prime}{}^{\Lambda} - 3\, {{\mathbb H}}^{\prime}{}^{\Lambda}\right) + i \left( - \, s \, {\mathbb P}^{}{}^{\Lambda} - 2\, {\mathbb Q}^{\prime}{}^{\Lambda} + 3\, s\, {\mathbb F}^{\prime}{}^{\Lambda} \right). \nonumber
\eea
Also, let us recall that the electric/magnetic components of the fluxes defined in (\ref{eq:chi-psi-Psi}), are expanded in the three-form basis as ${\rm Flux} = {\rm Flux}^\Lambda \, {\cal A}_\Lambda - {\rm Flux}_\Lambda \, {\cal B}^\Lambda,$ where ${\rm Flux} = \{\chi, \psi, \Psi\}$.

Subsequently, the full scalar potential can be expressed in just a few lines, and can be clubbed into two types of terms, namely $({\cal O}_1 \wedge \ast \ov {\cal O}_2)$ and $({\cal O}_1 \wedge \ov {\cal O}_2)$, which we present below,
\bea
\label{eq:symplectic-2}
& & V =  V_{({\cal O}_1 \wedge \ast \ov {\cal O}_2)} + V_{({\cal O}_1 \wedge \ov {\cal O}_2)},
\eea
where
\bea
\label{eq:symplectic-3}
& &  \hskip-1.5cm V_{({\cal O}_1 \wedge \ast \ov {\cal O}_2)} = -\frac{1}{4\,s\, {\cal V}^2}\,  \int_{X_6} \biggl[\chi \wedge \ast \ov\chi + {\widetilde\psi} \wedge \ast {\ov{\widetilde\psi}} + {\cal G}_{\alpha\beta} \, {\widetilde\Psi^\alpha} \wedge \ast \ov{\widetilde\Psi^\beta} \\
& & \hskip0.75cm + \frac{i}{2}\left(\widetilde{\chi} \wedge \ast  {\ov{\widetilde\psi}} - \ov{\widetilde{\chi}} \wedge \ast  {\widetilde\psi} \right) + \frac{i}{2}\left(\widetilde\Psi \wedge \ast \ov{\widetilde{\chi}} - \ov{\widetilde\Psi} \wedge \ast \widetilde{\chi} \right)  \biggr], \nonumber\\
& & \hskip-1.5cm V_{({\cal O}_1 \wedge \ov {\cal O}_2)} = -\frac{1}{4\,s\, {\cal V}^2}\,  \int_{X_6} \biggl[ (-i) \left(\chi \wedge \ov\chi + \chi \wedge \ov{\widetilde\chi} + 2 \, {\widetilde\psi} \wedge {\ov{\widetilde\psi}} + 2\,{\cal G}_{\alpha\beta} \,{\Psi^\alpha} \wedge \ov{\widetilde\Psi^\beta} \right)\nonumber\\
& & \hskip0.75cm + \left(\widetilde{\chi} \wedge  {\ov{\widetilde\psi}} + \ov{\widetilde{\chi}} \wedge {\widetilde\psi} \right) + \left(\widetilde\Psi \wedge \ov{\widetilde{\chi}} + \ov{\widetilde\Psi} \wedge \widetilde{\chi} \right) \biggr], \nonumber
\eea
where the so-called tilde fluxes for $\chi, \psi$ and $\Psi^\alpha$ are defined as below,
\bea
\label{eq:tilde-chi-psi-Psi}
& & \widetilde{\chi} = -\left({{\cal S}}^{\Sigma\Delta} \chi{}_\Delta + {{\cal S}}^\Sigma{}_\Delta \chi^{\Delta}\right) {\cal A}_\Sigma + \left({{\cal S}}_{\Sigma}{}^{\Delta} \chi{}_\Delta + {{\cal S}}_{\Sigma\Delta} \chi^{\Delta}\right) {\cal B}^\Sigma, \\
& & \widetilde{\psi} = -\left({{\cal S}}^{\Sigma\Delta} \psi{}_\Delta + {{\cal S}}^\Sigma{}_\Delta \psi^{\Delta}\right) {\cal A}_\Sigma + \left({{\cal S}}_{\Sigma}{}^{\Delta} \psi{}_\Delta + {{\cal S}}_{\Sigma\Delta} \psi^{\Delta}\right) {\cal B}^\Sigma, \nonumber\\
& & \widetilde{\Psi}^\alpha = -\left({{\cal S}}^{\Sigma\Delta} \Psi^\alpha{}_\Delta + {{\cal S}}^\Sigma{}_\Delta \Psi^\alpha{}^{\Delta}\right) {\cal A}_\Sigma + \left({{\cal S}}_{\Sigma}{}^{\Delta} \Psi^\alpha{}_\Delta + {{\cal S}}_{\Sigma\Delta} \Psi^\alpha{}^{\Delta}\right) {\cal B}^\Sigma \nonumber
\eea 
and using (\ref{coff}) and (\ref{coff2}) the ${\cal S}$ matrices are defined as
\begin{eqnarray}
\label{eq:S-matrices}
& &  {{\cal S}}^{\Lambda \Delta} = \left({\cal M}^{\Lambda}{}_{ \Sigma} \, {\cal L}^{\Sigma \Delta} + {\cal M}^{\Lambda \Sigma} \, {\cal L}_{\Sigma}{}^{ \Delta} \right), \\ 
& & {{\cal S}}^{\Lambda}{}_{ \Delta}  = - \left({\cal M}^{\Lambda}{}_{ \Sigma} \, {\cal L}^{\Sigma}{}_\Delta + {\cal M}^{\Lambda \Sigma} \, {\cal L}_{\Sigma \Delta}\right) + \delta^\Lambda{}_\Delta, \,\nonumber\\
& & {{\cal S}}_\Lambda^{\, \, \, \, \Delta} =\left({\cal M}_{\Lambda}{}_{ \Sigma} \, {\cal L}^{\Sigma \Delta} + {\cal M}_{\Lambda}{}^{ \Sigma} \, {\cal L}_{\Sigma}{}^{ \Delta} \right)  - \delta_\Lambda{}^\Delta\,, \nonumber\\
& &  {{\cal S}}_{\Lambda \Delta}\, =  - \left({\cal M}_{\Lambda \Sigma} \, {\cal L}^{\Sigma}{}_{ \Delta} + {\cal M}_{\Lambda}{}^{\Sigma} \, {\cal L}_{\Sigma \Delta} \right). \nonumber
\end{eqnarray}
In order to demonstrate the use of the mater formula in the symplectic proposal (\ref{eq:symplectic-2})-(\ref{eq:symplectic-3}), we get back to the toroidal type IIB model model based on $\mathbb T^6/(\mathbb Z_2 \times \mathbb Z_2)$ orientifold. As we have already observed for the previous formulation using the metric of the toroidal orbifold, the symplectic formulation (\ref{eq:master1}) also leads to a total of 76276 terms while being expressed in terms of the usual fluxes, however this number reduces to 10888 terms when the total scalar potential is expressed in terms of the axionic-fluxes (\ref{eq:AxionicFlux}), subsequently leading to the numerics about the number of terms in each of the 36 pieces as presented in Table \ref{tab_term-counting}. As a particular case, several scenarios can be considered by switching-off certain fluxes at a time.  Moreover, in light of recovering the results from our master formula (\ref{eq:symplectic-2})-(\ref{eq:symplectic-3}), we have a clear splitting of 10888 terms in the following manner,
\bea
& & \#\left(V_{({\cal O}_1 \wedge \ast \ov {\cal O}_2)}\right) = 5576, \qquad \qquad \#\left(V_{({\cal O}_1 \wedge \ov {\cal O}_2)}\right) = 5312.
\eea

%%%%%%%%%%%%%%%%%%%%%%%%%%%%%%%%%%%%%%%%%%%%%%%%%%%%%%%%%%%%%%%%%%%%%%%%%%%%%%%%%%%%%%%%%%
%%%%%%%%%%%%%%%%%%%%%%%%%%%%%%%%%%%%%%%%%%%%%%%%%%%%%%%%%%%%%%%%%%%%%%%%%%%%%%%%%%%%%%%%%%

%%%%%%%%%%%%%%%%%%%%%%%%%%%%%%%%%%%%%%%%%%%%%%%%%%%%%%%%%%%%%%%%%%%%%%%%%%%%%%%%%%%%%%%%%%
%%%%%%%%%%%%%%%%%%%%%%%%%%%%%%%%%%%%%%%%%%%%%%%%%%%%%%%%%%%%%%%%%%%%%%%%%%%%%%%%%%%%%%%%%%

\section{Summary and conclusions}

%%%%%%%%%%%%%%%%%%%%%%%%%%%%%%%%%%%%%%%%%%%%%%%%%%%%%%%%%%%%%%%%%%%%%%%%%%%%%%%%%%%%%%%%%%
%%%%%%%%%%%%%%%%%%%%%%%%%%%%%%%%%%%%%%%%%%%%%%%%%%%%%%%%%%%%%%%%%%%%%%%%%%%%%%%%%%%%%%%%%%

\noindent
In this work we presented a brief review about the various available formulations of non-geometric scalar potential arising from the generalized flux superpotential. The basic starting point is the four-dimensional type IIB flux superpotential which is induced by the standard three-form RR and NS-NS fluxes $(F, H)$ resulting in a cubic polynomial in complex structure moduli along with a linear dependence on the axio-dilaton modulus. A set of successive applications of S/T-dualities lead to the U-dual completion of the flux superpotential which has a cubic dependence for the complex structure moduli as well as the K\"ahler moduli, in addition to having a linear dependence on the axio-dilaton. We discussed multiple methods of arriving at the same $N = 1$ four-dimensional scalar potentials generated by the generalized flux superpotentials. These methods are:

\begin{itemize}

\item 
Method 1: This is the direct method of computing the scalar potential using the $N = 1$ formula (\ref{eq:Vtot}) for a given generic form of K\"ahler potential $K$ and the superpotenial $W$.

\item 
Method 2: In this method the scalar potential is calculated using the metric of the internal sixfold background. For the same reason this method is applicable for the cases where the metric of the compactified space is known, e.g. the case of toroidal models.

\item
Method 3: The notion of non-geometric fluxes has been motivated by the toroidal setup, and in order to understand/promote the idea to cases like Calabi Yau compactifications, one has to understand the insights of the effective scalar potential without the need of using the internal metric. For this purpose, the role of symplectic geometries gets very crucial \cite{Ceresole:1995ca, D'Auria:2007ay} and it turns out that the scalar potential can be equivalently expressed in terms of symplectic ingredients. We presented the master formula (\ref{eq:symplectic-2}) which is generically applicable to such cases.

\item 
Method 4: After writing all the relevant quantities in terms of symplectic ingredients (\ref{eq:symplectic-2}), one finds that although the formulation is quite generic and compact, making attempts for phenomenology or addressing any model building issues is still hard. This is because of the fact that the required symplectic quantities are implicit in terms of moduli dependence and it is hard to make  direct use of symplectic understanding at the level of scalar potential for the search of physical vacua. In this regard, a particular type IIB setting with the absence of prime fluxes has been studied in \cite{Shukla:2016hyy,Shukla:2019wfo} which have been extended for the most generic case recently in  \cite{Biswas:2024ewk}.  These formulations are such that one can directly {\rm read-off} the scalar potentials by using the topological data of the compactifying CY and its mirror threefold, and we hope that it can be useful for search physics vacua, including the de-Sitter ones in the vast landscape of the non-geometric fluxes.

\end{itemize}

\noindent
The main aim of formulating the scalar potential in various equivalent forms is to write down the same in a compact and concise way which can be useful and easy-to-handle for the phenomenological purposes such as moduli stabilization and the  search for de-Sitter vacua \cite{AbdusSalam:2024abc}. 

%%%%%%%%%%%%%%%%%%%%%%%%%%%%%%%%%%%%%%%%%%%%%%%%%%%%%%%%%%%%%%%%%%%%%%%%%%%%%%%%%%%%%%%%%%
%%%%%%%%%%%%%%%%%%%%%%%%%%%%%%%%%%%%%%%%%%%%%%%%%%%%%%%%%%%%%%%%%%%%%%%%%%%%%%%%%%%%%%%%%%

\subsection*{Acknowledgments}
\noindent
PS is thankful to the {\it Department of Science and Technology (DST), India} for the kind support.

%%%%%%%%%%%%%%%%%%%%%%%%%%%%%%%%%%%%%%%%%%%%%%%%%%%%%%%%%%%%%%%%%%%%%%%%%%%%%%%%%%%%%%%%%%
%%%%%%%%%%%%%%%%%%%%%%%%%%%%%%%%%%%%%%%%%%%%%%%%%%%%%%%%%%%%%%%%%%%%%%%%%%%%%%%%%%%%%%%%%%

%\newpage
%\bibliographystyle{JHEP}
%\bibliography{reference}

%\newpage
\bibliographystyle{utphys}
\bibliography{reference}

\end{document}